\documentclass[12pt, draftclsnofoot,onecolumn]{IEEEtran}
\usepackage[tight,footnotesize]{subfigure}
\usepackage{float}
\usepackage{hyperref}
\usepackage{bm}
\usepackage{cite}
\usepackage{graphicx}
\usepackage[ruled]{algorithm2e} 
\usepackage{algorithmic}
\usepackage{amssymb}
\usepackage{amsmath}
\usepackage{amsthm}
\usepackage{mathrsfs}
\usepackage{color}
\usepackage{booktabs}
\usepackage{bbm}
\usepackage{epstopdf}

\theoremstyle{definition}
\newtheorem{theorem}{Theorem}

\allowdisplaybreaks[1]
\allowdisplaybreaks

\newcommand{\tabincell}[2]{\begin{tabular}{@{}#1@{}}#2\end{tabular}} 

\begin{document}

\title{
Proximal Gradient-Based Unfolding
 for Massive Random Access in IoT Networks
\thanks{This paper was presented in part at the 
	 IEEE Global Communications Conference (Globecom), Madrid, Spain, 2021 \cite{zou2021learning}.}
}

\author{
  Yinan Zou,~\IEEEmembership{Graduate Student Member,~IEEE},~Yong~Zhou,~\IEEEmembership{Senior Member,~IEEE},  
  Xu Chen,~\IEEEmembership{Senior Member,~IEEE}, and Yonina C. Eldar,~\IEEEmembership{Fellow,~IEEE}
  	\thanks{Y. Zou and Y. Zhou are with the School of Information Science and Technology, ShanghaiTech University, Shanghai 201210, China (E-mail: \{zouyn,  zhouyong\}@shanghaitech.edu.cn). 
  	X. Chen is with the School of Computer Science and Engineering, Sun Yat-sen University, Guangzhou 510006, China (e-mail: chenxu35@mail.sysu.edu.cn). 
  	Yonina C. Eldar is with the Faculty of Math and CS, Weizmann Institute
  	of Science, Rehovot 7610001, Israel (email: yonina.eldar@weizmann.ac.il).}
}

\maketitle

\begin{abstract}

	Grant-free random access is an effective technology for enabling low-overhead and low-latency  massive access, where joint activity detection and channel estimation (JADCE) is a critical issue.
	Although  existing compressive sensing algorithms can be applied for JADCE,
	they usually fail to simultaneously harvest the following properties: effective sparsity inducing,  fast convergence, robust to different pilot sequences, and adaptive to time-varying networks.
	To this end, 
	we propose an unfolding   framework for JADCE based on the proximal gradient method.
	Specifically, we formulate the JADCE problem as a group-row-sparse matrix recovery problem and leverage a  minimax concave penalty rather than the widely-used $\ell_1$-norm to induce sparsity.
	We then develop a proximal gradient-based  unfolding neural network that parameterizes the algorithmic iterations.
	To improve convergence rate, we incorporate momentum into the unfolding neural network, and prove the accelerated convergence  theoretically.
	Based on the convergence analysis, we further develop an adaptive-tuning algorithm, which adjusts its parameters to  different  signal-to-noise ratio settings.
	Simulations show that the proposed  unfolding neural network achieves better  recovery performance, convergence rate, and adaptivity than current baselines.
	
\end{abstract}


\section{Introduction}

Massive machine-type communications (mMTC)  is expected to connect a massive number of 
  Internet of Things (IoT) devices 
 \cite{sharma2019toward}.
Because of the sporadic short-packet communication and massive connectivity,
 adopting the conventional grant-based random access strategy  to support  mMTC may lead to overwhelming signaling overhead, thereby introducing significant access latency. 
Grant-free random access has received extensive attention,
given its potential to enable low-latency and low-overhead massive access \cite{liu2018sparse}.
Specifically, without waiting for the grant, each IoT device directly transmits its data to the base station (BS) after sending a pilot sequence,
which significantly reduces the signaling overhead.
To fully exploit the advantages  of grant-free random access, 
it is essential to  achieve joint activity detection and channel estimation (JADCE) according to  the pilot sequences received at the BS.

 Because of the sporadic traffic of IoT devices and large antenna array at the BS, 
JADCE is usually modeled as different multiple measurement vector (MMV) compressive sensing (CS) problems \cite{eldar2012compressed,eldar2009robust,eldar2010block} and then  tackled by applying  sparse signal processing methods.
In particular, the JADCE problem can be formulated as group LASSO, which can be solved by the iterative shrinkage thresholding algorithm (ISTA) \cite{qin2013efficient,yuan2006model}.
Apart from ISTA, other optimization-based algorithms  \cite{jiang2018joint,shao2019dimension,shao2021exploiting,he2018compressive} have also been developed for JADCE.
 The authors in \cite{liu2018massive} proposed an  approximate message passing (AMP)-based algorithm for JADCE in massive multiple-input multiple-output (MIMO) systems.
AMP was further extended for activity detection in multi-cell  networks \cite{chen2019multi}.
In addition, the use of AMP for reconfigurable intelligent surface (RIS)-assisted  massive access systems was studied in  \cite{xia2021reconfigurable}.
Despite the aforementioned studies,  AMP-based algorithms may not converge in scenarios with either ill-conditioned or  non-Gaussian pilot sequences \cite{rangan2019vector,rangan2019convergence}.
Moreover, optimization-based methods often have slow convergence and high computation  complexity, and obtain sub-optimal solutions in practice, leading  to non-negligible performance gap to the optimal solution.


Deep learning (DL)
was emerged as a disruptive technique to tackle different optimization  problems in wireless networks \cite{eldar2022machine}, including  sparse signal  recovery.
 In order to enable model-driven learning design for sparse signal recovery, unfolding  iterative algorithms as recurrent neural networks (RNN) \cite{monga2021algorithm,scarlett2022theoretical} is an effective strategy.
  Different from the optimization-based methods that manually fix the parameters throughout the iterations, RNN adaptively tunes the 
 parameters in each unfolding layer according to the training  data, which accelerates convergence and leads to  performance improvement.
The authors in \cite{gregor2010learning,borgerding2017amp} proposed to unfold the generic ISTA and AMP into learned ISTA (LISTA) and learned AMP (LAMP), respectively.
The authors in \cite{chen2018theoretical,liu2019alista} simplified the LISTA structure by studying its theoretical properties and proved its linear convergence.
In \cite{shi2021algorithm}, a LISTA framework was developed for group sparsity. 
To improve recovery performance, \cite{cui2020jointly} considered an auto-encoder neural network to jointly design 
the pilot sequence matrix and recover sparse signal.
By exploiting the domain knowledge and channel structure, the authors in \cite{johnston2022model} proposed  DL-based approaches to aid the message passing algorithm.
An asynchronous grant-free random access system was studied in \cite{zhu2021deep}, where different LAMP-based structures were designed to balance the tradeoff between performance and complexity.
%
These studies \cite{shi2021algorithm,cui2020jointly,johnston2022model}
leveraged the widely-used $\ell_1$-norm as the sparsity-inducing penalty (SIP).

To further promote sparse solutions, 
a proximal operator method was unfolded as an RNN for non-convex SIP-regularized problems in \cite{yang2020learning}.
Though the scalar operator-based unfolding structure in \cite{yang2020learning} is effective for SMV problems, it does not consider the group-sparse structure that exists in the JADCE problem.
Furthermore, these DL-based methods \cite{shi2021algorithm,cui2020jointly,shao2021feature,johnston2022model,yang2020learning,zhu2021deep} are developed based on a common assumption that the training and test datasets share the same distribution, i.e.,  signal-to-noise ratio (SNR) and device active ratio  remain unchanged in the training and test stages.
However, in many practical IoT networks,   SNR and device active ratio are time-varying, which leads to a discrepancy  between the training and test datasets.
Hence, existing DL-based algorithms cannot be directly applied in such dynamic environments. 
An intuitive method to tackle this issue is to collect a new training dataset and re-train the neural network, which, however, incurs excessive communication and computation overhead  for data collection and training.
The authors in \cite{chen2021hyperparameter} proposed an adaptive scheme based on LISTA.
However, how to develop an adaptive method for JADCE problems with group-sparse channel matrix and non-convex SIP has not  been studied.

In this paper, we propose an adaptive unfolding neural network framework for JADCE based on a non-convex regularizer for group-sparsity,
which ensures robustness to non-Gaussian pilot sequences, achieves fast convergence with theoretical guarantees, and
 adapts to time-varying device active ratio and SNR.
As an effective approach to restrain oscillation and accelerate convergence,
we incorporate momentum into the
unfolding neural network.
The main contributions of this paper are summarized as follows:
\begin{itemize}
	\item We formulate the JADCE problem as a minimax concave penalty (MCP) regularized  group-row-sparse matrix recovery problem.
	To efficiently solve this challenging problem, we propose a light-weight unfolding 
	neural network, termed analytic learned proximal gradient method (ALPGM).
	\item 
	To further improve convergence rate, we incorporate  momentum into ALPGM and propose an accelerated variant  of ALPGM, termed ALPGM with momentum (ALPGM-MM).
	Theoretical analysis is conducted to characterize the convergence of ALPGM-MM.
	The theoretical result shows that ALPGM-MM has the no-false-positive property and enjoys a better convergence rate than ALISTA-GS in \cite{shi2021algorithm} under certain parameter settings.

	\item 
	Based on the convergence analysis,
	we further  propose an adaptive-tuning scheme, termed LPGM-AT, which adapts to the variation  of the device active ratio and SNR.
	The hyperparameters in LPGM-AT are optimized by grid search rather than back-propagation, which significantly reduces the computational complexity.
	The proposed LPGM-AT adaptively adjusts the network parameters according to the input data, and hence facilitates JADCE in time-varying IoT networks.

	\item 
	Simulations show that the proposed ALPGM and ALPGM-MM achieve better recovery performance than the baselines. 
	Moreover, benefiting from the momentum acceleration, the proposed ALPGM-MM  exhibits faster convergence rate than ALPGM.
	LPGM-AT significantly outperforms ALPGM and ALPGM-MM  on the test dataset 
	that differs from the training dataset in terms of device active ratio and SNR.

\end{itemize}

The remainder of this paper is organized as follows.
System model and problem formulation are described in Section II.
In Section III, we propose three  unfolding neural networks for tackling the JADCE problem.
We present simulation results in Section IV. 
Finally, the paper is concluded in Section V.

Notations: We denote $[N]=[ 1,\ldots,N]$.
We use $\mathbb{R}^N$ and $\mathbb{C}^N$ to denote the real and complex domains of dimension $N$, respectively,
$|S|$ denotes the cardinality of  set $S$ and $\text{supp}(\bm{x})$ is the support of  vector $\bm{x}=[x_1,\ldots,x_N]\in\mathbb{R}^N$. 
We denote the  sign function and the  generalized inverse of a matrix  as $\text{sign}(\cdot)$ and $\bm{X}^{\dagger}$, respectively.

%
%
%
%
%
%
%

\section{System Model and Problem Formulation}
\subsection{System Model}

In this paper, 
we consider a single-cell IoT network, which consists of 
$ N $ single-antenna IoT devices and one  $M$-antenna BS. 
Compared to the number of BS antennas, the  number of IoT devices is generally much larger, i.e., $N \gg M$. 
According to the principle of grant-free random access, each IoT device with sporadic traffic independently makes the transmission decision, 
and a small number  of IoT devices decide to transmit  in each transmission block. 
Specifically, the active devices, without the need to obtain a scheduling grant from the BS, send their pre-allocated 
pilot sequences along with their short-length data, while the inactive devices keep silent.
In any transmission block, we denote $a_n = 1$ if device $n$ is active, and $a_n = 0$ otherwise. 
The uplink channel response between IoT device $n$ and the BS is denoted as $ \bm {h}_n \in  \mathbb{C}^M$, which remains unchanged in each transmission block  and varies independently across different blocks and devices \cite{chen2018sparse}. 
With synchronized pilot transmissions from active devices,  the signal $ \bm{y}(\ell) \in \mathbb{C}^M $ received  at the BS is
\begin{equation}\label{eq01}
\bm{y}(\ell) = \sum_{n=1}^{N} \bm{h}_n a_n s_n(\ell) + \bm{z}(\ell),\quad  \ell = 1,\ldots, L,
\end{equation}
where $s_n(\ell) $ is the $ \ell $-th pilot symbol transmitted by device $n$, $L$ denotes the pilot length, and $ \bm{z}(\ell) \in \mathbb{C}^M $ denotes the additive white Gaussian noise (AWGN) vector with each entry following distribution $\mathcal{CN}(0,\sigma^2)$.
Compared to the device number, the pilot sequence length is generally much smaller, i.e., $L \ll N$, which makes it impractical for all devices to have  orthogonal sequences.
As a result, each device is assigned a non-orthogonal but unique sequence.

 \begin{figure}[tbp] 
 	\centering
 	\includegraphics[width=0.99\textwidth]{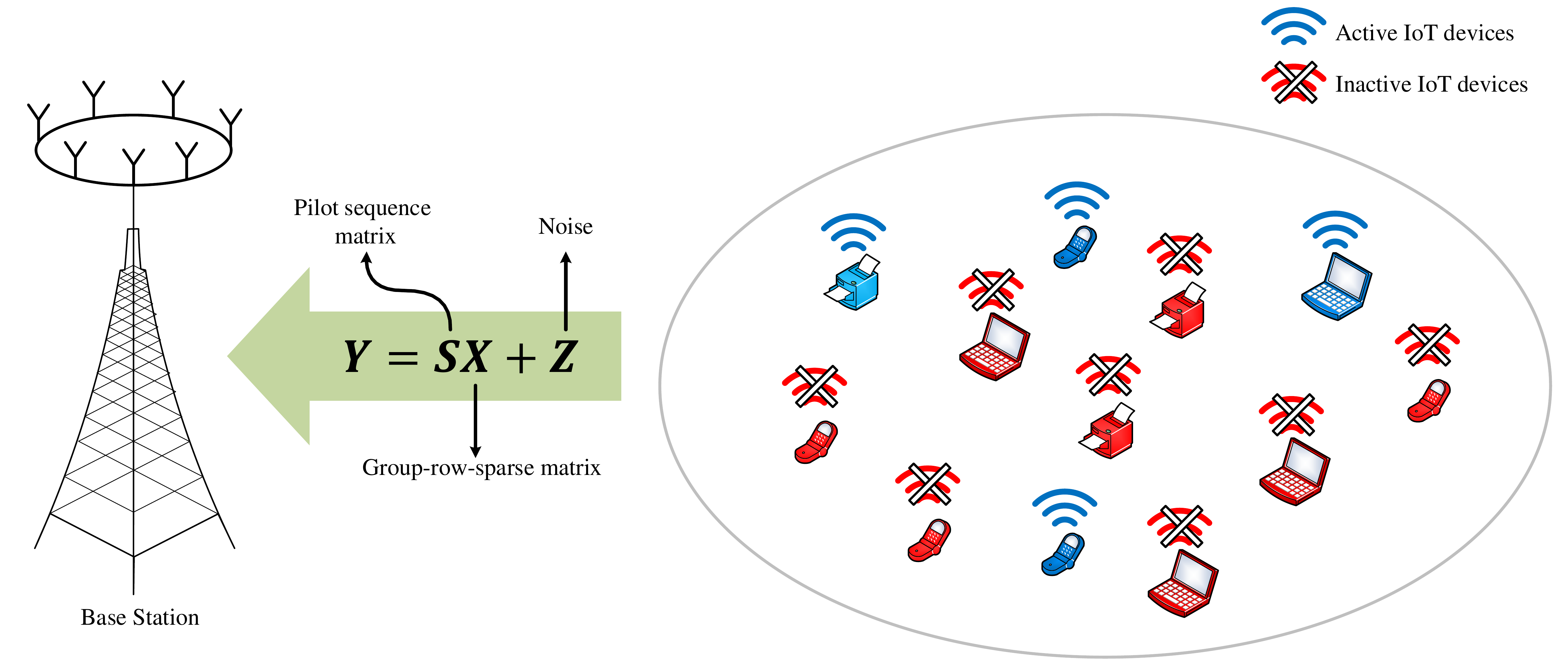}
 	\caption{An illustration of an IoT network that consists of massive devices with sporadic traffic.} 
 \end{figure}

By denoting  $ \bm{Y} = [\bm{y}(1), \ldots, \bm{y}(L)]^{\mathrm{T}} \in \mathbb{C}^{L \times M}$, 
$ \bm A = \operatorname{Diag}(a_1, \ldots, a_N) \in \mathbb{R} ^{N\times N} $, 
$ \bm{S}  =[\bm{s}(1), \ldots, \bm{s}(L)]^{\mathrm{T}} \in \mathbb{C}^{L \times N} $ with $ \bm{s}(\ell) = [s_1(\ell), \ldots,  s_N(\ell)]^{\mathrm{T}} \in \mathbb{C}^N$, $ \bm{H}  = [ \bm{h}_1, \ldots, \bm{h}_N]^{\mathrm{T}} \in \mathbb{C}^{N\times M}$,  and $ \bm{Z}  =[\bm{z}(1), \ldots, \bm{z}(L)]^{\mathrm{T}} \in \mathbb{C}^{L \times M} $, 
the received signal at the BS is rewritten  in matrix form as 
\begin{equation} \label{eq02}
	\bm{Y} = \bm{SAH} + \bm{Z}.
\end{equation}
Before decoding data, the BS conducts JADCE (i.e., recovering matrices $\bm{A}$ and $\bm{H}$) based on the received pilot signals.
Denoting $ \bm{X} = \bm{AH} \in \mathbb{C}^{N\times M} $, we rewrite (\ref{eq02}) as 
\begin{equation}\label{eq03}
\bm{Y}  =\bm{SX} + \bm{Z}. 
\end{equation}

\subsection{Problem Formulation}
 
Since the device activity matrix $\bm A$ is diagonal, we have $\bm{X}=[a_1 \bm{h}_1,\ldots,a_N \bm{h}_N]^{\text{T}}$. 
If device $n$ is inactive, then all entries of the $n$-th row of matrix $\bm{X}$ are zero. 
Thus, matrix $\bm{X} $ has the structure of group-sparsity in rows and all columns share the same support.
Achieving JADCE is equivalent to
recovering the row support of $\bm{X}$ and  the elements of  nonzero rows based on the noisy observation $\bm{Y}$ at the BS.
Such a matrix recovery problem is given by 
\begin{equation} \label{eq04}
\begin{aligned}
\mathcal{P} : \mathop{ \text{minimize}}_{\bm{X}\in\mathbb{C}^{N \times M} }
\frac{1}{2} \left\| \bm{Y-SX} \right\|_F^2 + \lambda G(\bm{X}),
\end{aligned}
\end{equation}
where $ \lambda > 0 $ is the regularization parameter, and $ G(\bm{X})  $ is an SIP term introduced to induce the group-row-sparsity of matrix $\bm X$.

In the following, (\ref{eq03}) is rewritten as its real-valued counterpart
\begin{align}\label{eq05}
\tilde{\bm{Y}}&=\tilde{\bm{S}}\tilde{\bm{X}}+\tilde{\bm{Z}}=
\begin{bmatrix} \mathcal{R}\left\{\bm{S}\right\} & -\mathcal{I}\left\{\bm{S}\right\} \\ \mathcal{I}\left\{\bm{S}\right\} & \mathcal{R}\left\{\bm{S}\right\} \end{bmatrix}
\begin{bmatrix} \mathcal{R}\left\{\bm{X}\right\}  \\  \mathcal{I}\left\{\bm{X}\right\} \end{bmatrix}
+
\begin{bmatrix} \mathcal{R}\left\{\bm{Z}\right\}  \\  \mathcal{I}\left\{\bm{Z}\right\} \end{bmatrix},
\end{align}
where $\mathcal{R}\{\cdot\}$ and $\mathcal{I}\{\cdot\}$ denote the real and imaginary parts of a complex matrix. 
Hence, problem $\mathcal{P}$ is rewritten as 
\begin{equation} \label{eq06}
\begin{aligned}
\mathcal{P}_r:\mathop{\text{minimize}}_{\tilde{\bm{X}}\in\mathbb{R}^{2N \times M}}
\frac{1}{2}\|\tilde{\bm{Y}}-\tilde{\bm{S}}\tilde{\bm{X}}\|_F^2+\lambda G(\tilde{\bm{X}}). 
\end{aligned}
\end{equation}
To induce a group-sparse  solution, the authors in \cite{jiang2020distributed,shi2021algorithm} adopted a  convex SIP in the form of $G(\tilde{\bm{X}})=\sum_{i=1}^{2N}\|\tilde{\bm{X}}_{i,:}\|_2$ (i.e., mixed $\ell_1/\ell_2$-norm), and reformulated
problem $\mathcal{P}_r$ as group LASSO \cite{yuan2006model}.
 Since MCP \cite{zhang2010nearly} induces further sparsity than the $\ell_1$-norm, we choose MCP as the SIP and rewrite problem $\mathcal{P}_r$ as the following group MCP problem \cite{breheny2015group}
\begin{align}\label{eq08}
	\text{Group MCP}:\quad \mathop{\text{minimize}}_{\tilde{\bm{X}}\in\mathbb{R}^{2N \times M}}
	\frac{1}{2}\|\tilde{\bm{Y}}-\tilde{\bm{S}}\tilde{\bm{X}}\|_F^2+\lambda \sum_{i=1}^{2N} g_{\eta}(\|\tilde{\bm{X}}_{i,:}\|_2),
\end{align}
where  
\begin{align}\label{eq07}
	g_{\eta}(z) = \left \{
	\begin{aligned}	
		& | z | - \eta z^2,
		&& \text{if} \, | z | \leq \frac{1}{2\eta}, \\
		& \frac{1}{4\eta}, 
		&& \text{if} \, | z | > \frac{1}{2\eta}. 
	\end{aligned}
	\right. 
\end{align}

\subsection{Conventional Proximal Gradient Method}
For group MCP, we apply the following iterative proximal gradient method (PGM) to recover real-valued matrix $\tilde{\bm{X}}$
\begin{equation} \label{eq09}
\begin{aligned}
\tilde{\bm{X}}^{k+1}=P_{\lambda \gamma_k,f_{\eta_k}}\left(\tilde{\bm{X}}^k + \gamma_k\tilde{\bm{S}}^{\text{T}}(\tilde{\bm{Y}} - \tilde{\bm{S}} \tilde{\bm{X}}^k   ) \right),
\end{aligned}
\end{equation}
where $\gamma_k$ denotes the step-size and $\tilde{\bm{X}}^{k}$ is an estimation of  $\tilde{\bm{X}}$ at iteration $k$.
The multivariate proximal operator $P_{\lambda \gamma_k,f_{\eta_k}}(\cdot)$ is given by
\begin{align}\label{eq10}
	P_{\theta_k,f_{\eta_k}}(\tilde{\bm{X}}_{i,:})
	=\arg\min_{\tilde{\bm{U}}_{i,:}} 
	\frac{1}{2} \| \tilde{\bm{U}}_{i,:} - \tilde{\bm{X}}_{i,:} \|_2^2 + f_{\eta_k}(\tilde{\bm{U}}_{i,:} ),
\end{align}
with $f_{\eta_k}(\tilde{\bm{U}}_{i,:} )=\theta_k  g_{\eta_k}(\|\tilde{\bm{U}}_{i,:}\|_2)$ and  $\theta_{k} = \lambda \gamma_k$.
The univariate proximal operator  can be written as
$
	\hat{P}_{\theta_k,f_{\eta_k}}(x)
	=\arg\min_{u} 
	\frac{1}{2} ( u - x )^2 + \hat{f}_{\eta_k}(u),
$
where $\hat{f}_{\eta_k}(u)=\theta_k g_{\eta_k}(u)$ \cite{parikh2014proximal}.
To have a well-defined minimum \cite{zou2021learning},  we should have $\eta_k<\frac{1}{2\theta_k}$, which yields 
\begin{align}\label{eq12}
	\hat{P}_{\theta_k,f_{\eta_k}}(x) = \left \{
	\begin{aligned}
		& 0, 
		&&\text{if} \, |x| \leq \theta_k , \\
		& \frac{x-\theta_k \text{sign}(x)}{1-2\theta_k {\eta_k}}, &&\text{if} \, \theta_k < | x | \leq \frac{1}{2\eta_k}, \\
		& x,      
		&&\text{if} \, |x| > \frac{1}{2\eta_k}.
	\end{aligned}
	\right. 
\end{align}
Based on \cite[Theorem 6.18]{beck2017first} and \eqref{eq12},  we obtain
\begin{align}\label{eq13}
	P_{\theta_k,f_{\eta_k}}(\tilde{\bm{X}}_{i,:}) = 
	\left \{
	\begin{aligned}
		& \hat{P}_{\theta_k,f_{\eta_k}}(\|\tilde{\bm{X}}_{i,:}\|_2)
		\frac{\tilde{\bm{X}}_{i,:}}{\|\tilde{\bm{X}}_{i,:}\|_2},
		&& \text{if} \, \tilde{\bm{X}}_{i,:}\neq \bm{0},
		\\& \bm{0},
		&& \text{otherwise}.
	\end{aligned}
	\right. 
\end{align}

The resulting PGM can solve problem \eqref{eq08} \cite{breheny2011coordinate}.
However, it has several limitations. 
First, PGM achieves  sublinear convergence rate and usually takes many iterations to converge. 
In time-varying IoT networks, the variations of device active ratio and SNR cause PGM to re-execute, which incurs a high computational complexity.
Second, an inappropriate choice of the regularization parameter $\lambda$ may severely degrade  performance  of PGM. 
Third, the values of the step-size $\gamma_k$ and parameter $\eta_k$
 influence the convergence rate, and are generally tricky to choose.
To tackle these limitations,  we propose an unfolding neural network framework to improve recovery performance and accelerate the convergence  by learning key parameters $\lambda$, $\gamma_k$, and $\eta_k$.

\section{Proposed Unfolding Framework}
This section proposes  an unfolding framework that tackles  the matrix recovery problem by unfolding  the conventional PGM discussed in Section II-C.

\subsection{ALPGM}

Following the idea of algorithm unfolding, we unfold the iteration in \eqref{eq09} as an RNN.  
By treating $\tilde{\bm{X}}^k$ and $\tilde{\bm{X}}^{k+1}$ as the input and output of the activation function $P_{\lambda \gamma_k,f_{\eta_k}}(\cdot)$, respectively,  (\ref{eq09}) can be mapped to a one-layer neural network. 
Therefore, the $K$ iterations are implemented by  a $K$-layer RNN, where each  neural network layer corresponds  to a specific iteration  of PGM.
Motivated by \cite{chen2018theoretical,liu2019alista}, we replace $\tilde{\bm{S}}^{\text{T}}$ by matrix $\bm{B}$ which can be obtained 
before the training phase  via solving the following optimization problem
\begin{align} \label{eq16}
	\mathop{\text{minimize}}_{\bm{B}\in \mathbb{R}^{2N\times2L}} \quad & \|   \bm{B}\tilde{\bm{S}}
	\|_F^2 \\
	\text{subject to} \quad &\bm{B}_{i,:} \tilde{\bm{S}}_{:,i} = 1, \, \forall i \in [2N]. 
\end{align}
We utilize the projected gradient descend (PGD) method to solve problem \eqref{eq16} \cite{liu2019alista}.
The unfolding neural network termed ALPGM is thus given by
\begin{equation} \label{eq18}
	\begin{aligned}
		\tilde{\bm{X}}^{k+1}\!=\!P_{\theta_k,f_{\eta_k}}\left(\tilde{\bm{X}}^k\! +\! \gamma_k \bm{B}(  \tilde{\bm{Y}}\! -\! \tilde{\bm{S}} \tilde{\bm{X}}^k ) \right),\quad k =0,\ldots,K-1,
	\end{aligned}
\end{equation}
where $\theta_k = \lambda \gamma_k$ is the thresholding parameter of layer $k$. 
The trainable parameters are $ \bm{\Theta} = \{  \gamma_k, \theta_k, \eta_k \}_{k=0}^{K-1} $.   
The proposed ALPGM is shown in Fig. 2. 

\begin{figure}[tbp] 
	\centering
	\includegraphics[width=0.98\linewidth]{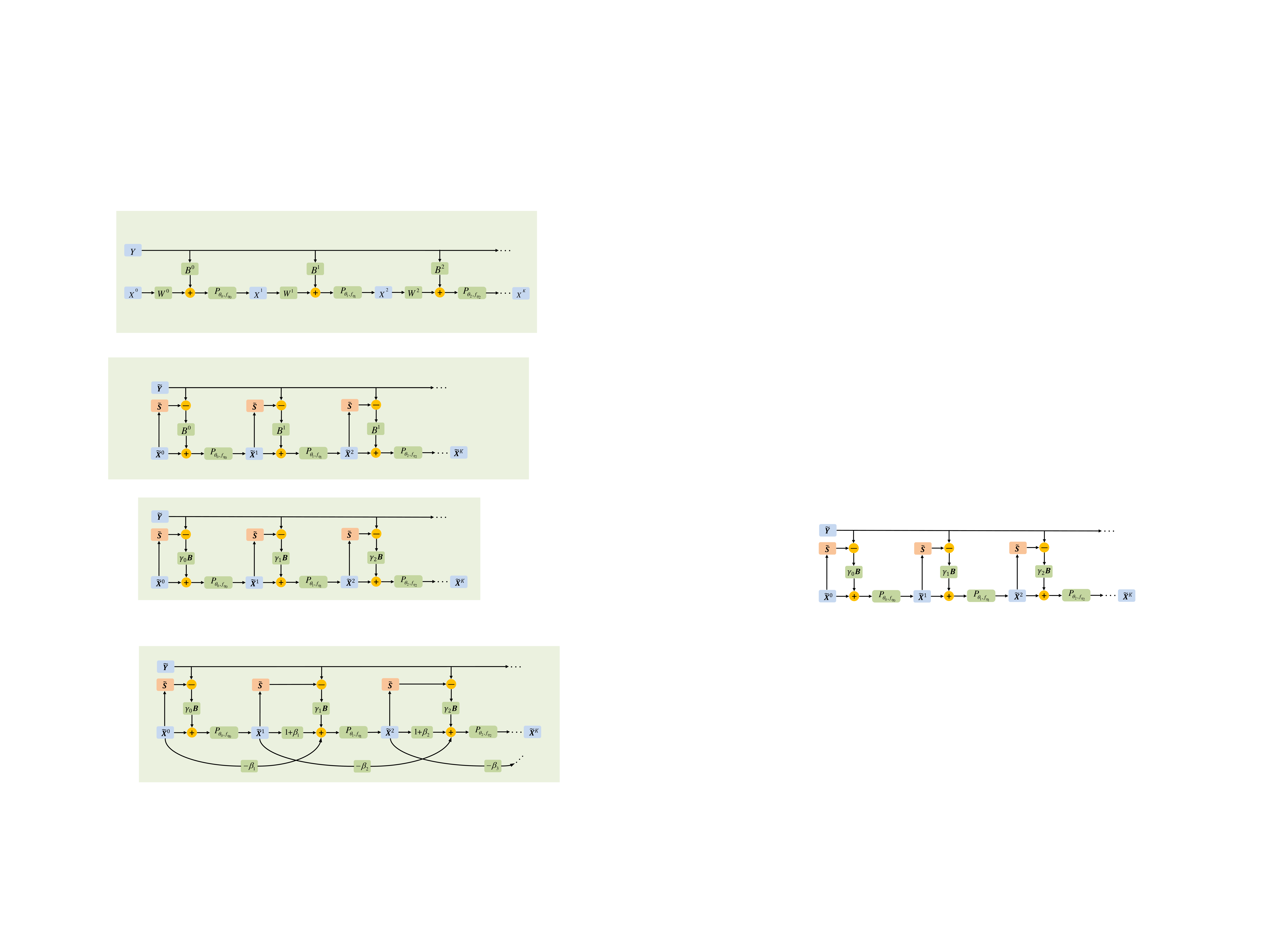}
	\caption{An illustration of the proposed ALPGM, where  $ \{ \gamma_k, \theta_k, \eta_k \}_{k=0}^{K-1}$ are trainable parameters.}  
\end{figure}

	We note that an intuitive method to unfold \eqref{eq09} is to fix $\tilde{\bm{S}}^{\text{T}}$ and then directly learn $\{\gamma_k,\theta_{k},\eta_{k}\}$.
	Another method  is to replace $\gamma_k\tilde{\bm{S}}^{\text{T}}$ by $\bm{B}^k$ and then learn $\{\bm{B}^k,\theta_{k},\eta_{k}\}$.
	These two methods achieve poorer  recovery performance and slower convergence rate than our proposed ALPGM, as will be shown in Section IV-A.

%
%
%


\subsection{ALPGM-MM}

For vanilla gradient descent,  the gradient may not always point towards the minimum, which results in an oscillating update path and slow convergence.
One solution is to utilize momentum  to mitigate oscillations and speed up convergence \cite{polyak1964some}.
Hence, we propose ALGPM-MM
where we introduce a momentum term relating to $\tilde{\bm{X}}^{k-1}$ into the update of $\tilde{\bm{X}}^{k+1}$ in ALGPM, i.e.,
\begin{align}\label{eq19}
	\tilde{\bm{X}}^{k+1}\!=
	\left\{
	\begin{aligned}
		&P_{\theta_k,f_{\eta_k}}\left(\tilde{\bm{X}}^k + \gamma_k \bm{B}(  \tilde{\bm{Y}} - \tilde{\bm{S}} \tilde{\bm{X}}^k ) \right), \, &&\text{if}\,k = 0,
		\\&	P_{\theta_k,f_{\eta_k}}\left(\tilde{\bm{X}}^k+\gamma_k \bm{B}(  \tilde{\bm{Y}} - \tilde{\bm{S}} \tilde{\bm{X}}^k )
		+ \beta_k(\tilde{\bm{X}}^k-\tilde{\bm{X}}^{k-1}) \right),\,  &&\text{if}\,k =1,\ldots,K-1,
	\end{aligned}
	\right.
\end{align}
where $\beta_k$ is the momentum parameter.
The trainable parameters are $ \{ \gamma_k, \theta_k, \eta_k \}_{k=0}^{K-1}$ and $\{\beta_k\}_{k=1}^{K-1}$. 
The proposed ALPGM-MM is shown in Fig. 3.

\begin{figure}[tbp] 
	\centering
	\includegraphics[width=0.98\linewidth]{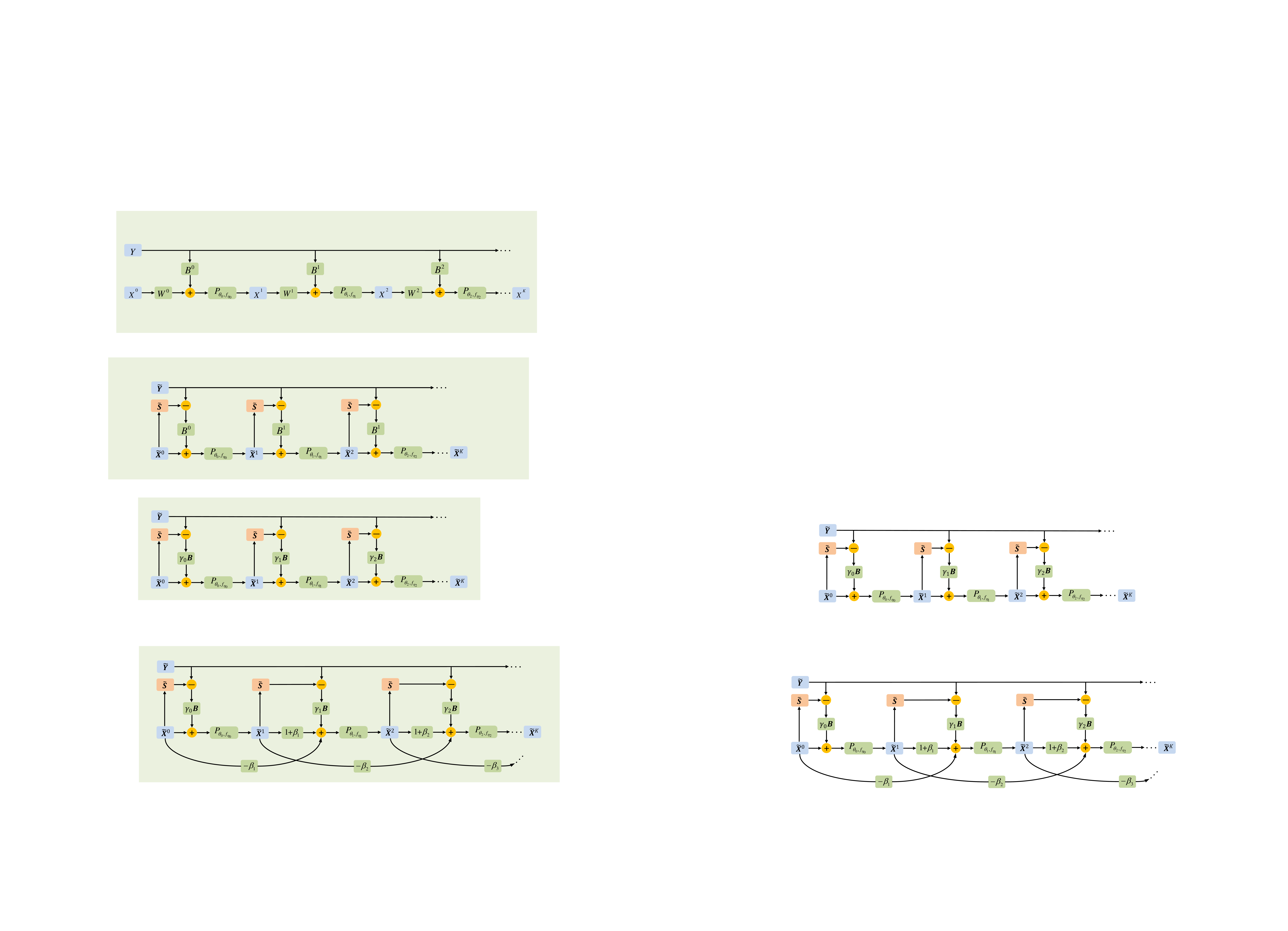}
	\caption{An illustration of the proposed ALPGM-MM, where  $ \{ \gamma_k, \theta_k, \eta_k \}_{k=0}^{K-1}$ and $\{\beta_k\}_{k=1}^{K-1}$ are trainable parameters. } 
\end{figure}

Since the update of $\tilde{\bm{X}}^{k+1}$ is dependent upon   $\tilde{\bm{X}}^{k}$ and $\tilde{\bm{X}}^{k-1}$,
it is difficult to directly analyze the convergence of ALPGM-MM.
Besides, the multivariate proximal operator with respect to MCP also brings a critical challenge for convergence analysis.          
For tractability of  the convergence analysis of ALPGM-MM,  the following problem replaces \eqref{eq16} to ensure  that the matrix $\bm{B}\tilde{\bm{S}}$ is symmetric.
By defining $\bm{B}= ((\bm{G}^\text{T}\bm{G})\tilde{\bm{S}})^\text{T}\in\mathbb{R}^{2N\times2L} $ with $\bm{G}\in\mathbb{R}^{2L\times2L}$, \eqref{eq16} can be written as
\begin{align}
	\mathop{\text{minimize}}_{\bm{G}\in \mathbb{R}^{2L\times2L}}  \quad &\|   \tilde{\bm{S}}^\text{T}\bm{G}^\text{T}\bm{G}\tilde{\bm{S}}-\bm{I}
	\|_F^2  \label{eq20}
	\\
	\text{subject to} \quad & (\tilde{\bm{S}}^\text{T}\bm{G}^\text{T}\bm{G}\tilde{\bm{S}})_{i,i} = 1,\,\forall i \in [2N]. \label{eq21}
\end{align}
Then, we define an auxiliary matrix $\bm{D}=\bm{G}\tilde{\bm{S}} \in \mathbb{R}^{2L\times2N}$ and reformulate \eqref{eq20} as 
\begin{align}\label{eq22}
	\mathop{\text{minimize}}_{\bm{G}\in \mathbb{R}^{2L\times2L}, \atop \bm{D}\in \mathbb{R}^{2L\times2N}}  \quad &\|   \bm{D}^\text{T}\bm{D}-\bm{I}
	\|_F^2
	 + \tau \|\bm{D}-\bm{G}\tilde{\bm{S}}\|_F^2
	 \\
	\text{subject to} \quad & 
	(\bm{D}^\text{T}\bm{D})_{i,i}=1,\,\forall i \in [2N],
\end{align}
where $\tau >0$ denotes the  regularization parameter.
We can also adopt  the PGD method to solve this problem \cite{ lu2018optimized}.
Through the above reformulation, matrix $\bm{B}\tilde{\bm{S}}
$ is guaranteed to be a positive semidefinite matrix.
In summary, before the training phase of ALPGM-MM, we solve problem (\ref{eq22}) to obtain $\bm{G}$, and then obtain $\bm{B}= ((\bm{G}^\text{T}\bm{G})\tilde{\bm{S}})^\text{T}$.

In Theorem 1, we show that ALPGM-MM has the no-false-positive property and achieves a faster convergence rate than ALISTA-GS in \cite{shi2021algorithm}.
We denote $\psi(\tilde{\bm{X}})=[ \|\tilde{\bm{X}}_{1,:} \|_2 ,\ldots,\|\tilde{\bm{X}}_{2N,:} \|_2  ]^\text{T}$
and define the mutual coherence of $\bm{D}$ as 
$
\phi \triangleq \max_{i \neq j} |\bm{D}_{:,i}^\text{T}\bm{D}_{:,j}|
$.
As in \cite{shi2021algorithm,chen2018theoretical,liu2019alista,chen2021hyperparameter,yang2020learning},  signal $\tilde{\bm{X}}^*$ and noise $\tilde{\bm{Z}}$ are assumed to belong to the set
$\mathcal{X}(\underline{\mu_x},\mu_x,s,\epsilon) \triangleq \{(\tilde{\bm{X}}^*,\tilde{\bm{Z}}) 
\,|\, 0<\underline{\mu_x}\leq\| \tilde{ \bm{X}}^*_{i,:} \|_2 \leq \mu_x, \forall\,i\in S, |S| \leq s,
\|\tilde{\bm{Z}} \|_F \leq \epsilon
\}$, where $\text{supp}(\psi(\tilde{\bm{X}}^{*}))$ is denoted as $S$.

\begin{theorem}
	\emph{
	For  ALPGM-MM, we denote the input as $\tilde{\bm{Y}}=\tilde{\bm{S}}\tilde{\bm{X}}^*+\tilde{\bm{Z}}$ and $\tilde{\bm{X}}^0=\bm{0}$, the output as $\{\tilde{\bm{X}}^k\}_{k=1}^{\infty}$,
	and 
	 $\|\bm{X}\|_{2,1}=\sum_n\|\bm{X}_{n,:}\|_2$. 
	If   $c_{\phi s}\triangleq(2s-1)\phi < 1$, 
	$\|\bm{B}\|_{2,1}\leq \mu_{B}$, 
	and the parameters $\{\theta_{k},\eta_k,\gamma_k,\beta_k\}$ satisfy
	\begin{align}
		&   \phi \mathop{\text{sup}}_{(\tilde{\bm{X}}^*,\tilde{\bm{Z}})\in\mathcal{X}(\underline{\mu_x},\mu_x,s,\epsilon)  }  \| \tilde{\bm{X}}^k  - \tilde{ \bm{X}}^* \|_{2,1} + \mu_{B} \epsilon 
		= \theta_k \leq \frac{1}{2 \eta_k}, \quad \forall\,k, \label{eq28}
		\\&\gamma_k=1, \quad \forall\,k, \label{eq29}
		\\&\beta_k \rightarrow 
		\frac{1}{2s}\bigg(1-\sqrt{1-c_{\phi s}}\bigg)^2,
		\quad \text{as}\, k \rightarrow \infty, \label{eq30}
		\\& \eta_k \rightarrow \frac{1}{2\theta_{k}}, 
		\quad \text{as}\, k \rightarrow \infty,\label{eq31}
	\end{align}
	then 
	the sequence of iterations in \eqref{eq19} satisfies
	\begin{align}
	\text{supp}(\psi (\tilde{ \bm{X}}^{k})) \subseteq S , \quad \forall\, k,
	\end{align}
	and
	\begin{align}
		\| \tilde{\bm{X}}^k  - \tilde{ \bm{X}}^* \|_{F}
		\leq C_0 \prod_{t=1}^{k}
		c^t 
		+  
		\frac{(1+s) \mu_{B} \epsilon}{1-c_{\phi s}},  \quad \forall\, k,
	\end{align}
	where $C_0>0$ is a constant and  $c^k$ satisfies
	\begin{align}
		&0<c^k \leq c_{\phi s} < 1, \quad \forall\, k,
		\\&0<c^k \leq 1-\sqrt{1-c_{\phi s}},
		\quad 
		\forall\, k > \bigg\lceil \frac{\log(\underline{\mu_x})-\log(6C_0)}{\log(c_{\phi s})}\bigg\rceil +2.
	\end{align} } 
\end{theorem}
\proof  See Appendix A.  \qedhere


%
%

	According to Theorem 1,  as long as the parameters satisfy \eqref{eq28}-\eqref{eq31},
	the index set of the rows containing
	non-zero elements of $\psi (\tilde{ \bm{X}}^{k})$ belongs to that of the ground truth.
	Based on the no-false-positive property, we prove that $\tilde{\bm{X}}^k$  converges to the vicinity of the ground truth $\tilde{\bm{X}}^*$, i.e., $\tilde{\bm{X}}^k$ is close to $\tilde{\bm{X}}^*$, which is group-row-sparse.
	Theorem 1 also demonstrates that ALPGM-MM achieves a linear convergence rate in a noisy scenario.
	Although the convergence rate is  linear, the convergence rate of ALPGM-MM (i.e., $c_{\phi s}$) is better than the convergence rate in \cite{zou2021learning} (i.e., $\phi s (\sqrt{M}+1) -\phi$).
	This is because the multivariate proximal operator exploits the group-row-sparsity property, which accelerates the convergence. 
	In addition, since $1-\sqrt{1-c_{\phi s}}< c_{\phi s}$ when $c_{\phi s}<1$, the convergence performance  of ALPGM-MM is better than that of ALISTA-GS in \cite{shi2021algorithm} under the same setting because the momentum term provides convergent acceleration.

	As ALPGM shares a similar network with ALPGM-MM except the momentum part, the convergence analysis of ALPGM-MM can be reduced to that of ALPGM by removing the momentum part.
	Through some modifications of the proof of Theorem 1, one prove that ALPGM  also achieves linear convergence rate.

%
%
%
%
%
%
%
%
%
%

	From a theoretical perspective, Theorem 1 verifies the validity  of  ALPGM-MM under certain conditions of parameters $\{\theta_{k},\eta_k,\gamma_k,\beta_k\}$.
	This assumption is only made for tractability of the analysis.
	Although the parameters learned by back-propagation may not necessarily satisfy the conditions in Theorem 1,
	ALPGM-MM with learned parameters still  exhibits excellent performance and fast convergence  in practice, as we show in Section IV-B.

%
%
%

\subsection{LPGM-AT}
Most DL-based approaches including the proposed ALPGM and ALPGM-MM rely on the assumption that the  SNR and the device active ratio remain the same during the training and test stages.
As a result, they may not work well in dynamic IoT networks, where the SNR and device active ratio are time-varying.
To  tackle this problem, we further develop an  adaptive-tuning algorithm, termed LPGM-AT, for dynamic IoT networks.

	In ALPGM-MM, $ \{ \gamma_k, \theta_k, \eta_k , \beta_k\}$ are regarded as the trainable parameters, and optimized by back-propagation on the training dataset.
	Thus, the optimized parameters entirely depend on training data and are applicable for test data that follows  the same distribution with  training data.
	The drawback is that a minor discrepancy  between training and test data distributions may incur severe performance degradation.
	To address this issue and achieve algorithmic robustness, we turn our attention to optimize the parameters according to  $\tilde{\bm{X}}^k$ and $\tilde{\bm{Y}}$.
	We design the adaptive-tuning update formulas of $ \{ \gamma_k, \theta_k, \eta_k , \beta_k\}$ as follows 
	\begin{align}
		&   \theta_{k} = c_{\theta} \|\tilde{\bm{S}}^{\dagger}(\tilde{\bm{S}}\tilde{\bm{X}}^k-\tilde{\bm{Y}})\|_{2,1},\quad k=0,\ldots,K-1, \label{eq32}
		\\&\gamma_k=1, \quad \forall\,k, \label{eq32.5}
		\\&\beta_k = c_{\beta}  \|\psi(\tilde{\bm{X}}^k)\|_0, 
		\quad k=1,\ldots,K-1,  \label{eq33}
		\\& \eta_k = \frac{1}{c_{\eta}\|\psi(\tilde{\bm{X}}^k)\|_0\theta_{k}},
		\quad k=0,\ldots,K-1, \label{eq35}
	\end{align}
	where $c_{\theta}>0, c_{\beta}>0, c_{\eta}>0$ are tunable hyperparameters.
	
	In the following we motivate the choices of (\ref{eq32})-(\ref{eq35}).  Starting with (\ref{eq32}), since $\tilde{\bm{Y}}=\tilde{\bm{S}}\tilde{\bm{X}}^*+\tilde{\bm{Z}}$, we obtain $\tilde{\bm{S}}^{\dagger}\tilde{\bm{Y}}=\tilde{\bm{S}}^{\dagger}\tilde{\bm{S}}\tilde{\bm{X}}^*+\tilde{\bm{S}}^{\dagger}\tilde{\bm{Z}}$, where $\tilde{\bm{S}}^{\dagger}$ is the  generalized inverse of  $\tilde{\bm{S}}$.
	Through adding $\tilde{\bm{S}}^{\dagger}\tilde{\bm{S}}\tilde{\bm{X}}^k$ and taking the norm on both sides,
	we obtain $\|\tilde{\bm{S}}^{\dagger}\tilde{\bm{S}}\tilde{\bm{X}}^k-\tilde{\bm{S}}^{\dagger}\tilde{\bm{Y}}\|_{2,1}=\|\tilde{\bm{S}}^{\dagger}\tilde{\bm{S}}(\tilde{\bm{X}}^k-\tilde{\bm{X}}^*)-\tilde{\bm{S}}^{\dagger}\tilde{\bm{Z}}\|_{2,1}$.
	We use $\|\tilde{\bm{S}}^{\dagger}(\tilde{\bm{S}}\tilde{\bm{X}}^k-\tilde{\bm{Y}})\|_{2,1}$ to approximate (\ref{eq28}) because $\|\tilde{\bm{S}}^{\dagger}\tilde{\bm{S}}(\tilde{\bm{X}}^k-\tilde{\bm{X}}^*)-	\tilde{\bm{S}}^{\dagger}\tilde{\bm{Z}}\|_{2,1}
	\approx 
	\mathcal{O}(\|\tilde{\bm{X}}^k-\tilde{\bm{X}}^*\|_{2,1})
	+ \mathcal{O}(\epsilon)$.
	In (\ref{eq28}), the thresholding parameter $\theta_{k}$ relies on  $\tilde{\bm{X}}^k$ and ground truth $\tilde{\bm{X}}^*$.
	By comparing (\ref{eq32}) with (\ref{eq28}), we observe that the thresholding parameter $\theta_{k}$ only depends on  $\tilde{\bm{X}}^k$ and $\bm{Y}$, and does not need the prior knowledge of $\tilde{\bm{X}}^*$.
	Second, for step-size parameter $\gamma_k$, we set $\gamma_k=1, k=0,\ldots,K-1$ according to \eqref{eq29}.
	Third, in \eqref{eq30},  the momentum parameter $\beta_k$ approaches 
	$ \frac{1}{2s}(1-\sqrt{1-c_{\phi s}})^2$ with $c_{\phi s}=(2s-1)\phi$ when $k$ approaches  infinity.
	Note that $\frac{1}{2s}(1-\sqrt{1-c_{\phi s}})^2$ is a monotonic increasing function of $s$ when $s> 1$.
	By following the same idea of getting rid of the dependence on $\tilde{\bm{X}}^*$, 
	we utilize $\|\psi(\tilde{\bm{X}}^k)\|_0$ to approximate $\frac{1}{2s}(1-\sqrt{1-c_{\phi s}})^2$,
	 because $\tilde{\bm{X}}^k$ converges to $\tilde{\bm{X}}^*$ while   $\|\psi(\tilde{\bm{X}}^k)\|_0$ approaches  $\|\psi(\tilde{\bm{X}}^*)\|_0$.
 Finally, by considering the coupling relationship between $\eta_k$ and $\theta_k$ (i.e., $2\theta_k\eta_k<1$), we design the adaptive-tuning update formula of  parameter $\eta_k$ as (\ref{eq35}).
We use grid search to find the best hyperparameters (i.e., $c_{\theta}$, $c_{\beta}$, and $c_{\eta}$) instead of back-propagation in the training phase.
Specifically, we execute the algorithm on the training dataset with a series of hyperparameter combinations and choose the hyperparameter combination that achieves the best performance.
LPGM-AT only needs to optimize three hyperparameters, which significantly reduces the training complexity.
Although DL  can also be leveraged for optimizing the three hyperparameters, it entails a much higher computational complexity than grid search.

The values of hyperparameters $\{c_{\theta}, c_{\beta}, c_{\eta}\}$ are determined in the training phase. 
Once the training phase ends, the hyperparameters are fixed, and directly applied to the test datasets.
According to (\ref{eq32})-(\ref{eq35}), parameters $\{\theta_k,\beta_k,\eta_k\}$ rely on hyperparameters $\{c_{\theta},c_{\beta},c_{\eta}\}$, $\tilde{\bm{X}}^k$, and $\tilde{\bm{Y}}$.
As the hyperparameters are fixed in the test phase, parameters $\{\theta_k,\beta_k,\eta_k\}$ only depend on  $\tilde{\bm{X}}^k$ and $\tilde{\bm{Y}}$.
For different distributions of the test dataset, 
parameters $\{\theta_k,\beta_k,\eta_k\}$ vary with $\tilde{\bm{X}}^k$ and $\tilde{\bm{Y}}$.
Thus, our proposed LPGM-AT is self-adaptive for different test datasets.
If the test dataset shares the same distribution with the training dataset, LPGM-AT achieves the same performance on both datasets.
If the distribution of the test dataset differs from that of the training dataset, then LPGM-AT  adapts to the unknown distribution of the test dataset.

%
%
%


%
%
%
%

\subsection{Training and Testing Strategies}
\subsubsection{ALPGM and ALPGM-MM}
For these two neural networks, we adopt  supervised learning
based on training set $ \{ \tilde{\bm{X}}_i^* , \tilde{\bm{Y}}_i \} _{i=1}^T $, where $\tilde{\bm{Y}}_i$ is the data, $ \tilde{\bm{X}}_i^*$ is the corresponding label, and $T$ is the size of the training set. 
We denote the output of $K$-layer RNN as $ \tilde{\bm{X}}^{K}(\bm{\Theta },\tilde{\bm{Y}}_i,\tilde{\bm{X}}^0) $, where  $ \tilde{\bm{Y}_i } $ and $ \tilde{\bm{X}}^0$ are the inputs of the $K$-layer RNN.
Given $ \{ \tilde{\bm{X}}_i^* , \tilde{\bm{Y}}_i \} _{i=1}^T $, 
we obtain the parameters of  $K$-layer RNN via solving the following problem
\begin{equation} 
\begin{aligned}
\bm{\Theta}^* = \text{arg} \mathop{\text{min}}_{\bm{\Theta }} \sum_{i=1}^{T}\left\| \tilde{\bm{X}}^{K}(\bm{\Theta },\tilde{\bm{Y}}_i,\tilde{\bm{X}}^0)-\tilde{\bm{X}}^*_i  \right\|_F^2.
\end{aligned}
\end{equation}

To  avoid converging to a local minimum,  the network parameters are trained layer-by-layer  \cite{borgerding2017amp}. 
We take the training of the parameters of 
layer $k$, denoted as $\bm{\Theta}_{k-1}$, as an example, which is performed  after the parameters of the first $(k-1)$ layers, denoted as $\bm{\Theta}_{0:k-2}$, are trained. To optimize $\bm{\Theta}_{k-1}$, we need to solve problem
\begin{align}
\mathop{\text{min}}_{\bm{\Theta }_{k-1}}
\sum_{i=1}^{T}\| \tilde{\bm{X}}^{k}(\bm{\Theta }_{0:k-1},\tilde{\bm{Y}}_i,
\tilde{\bm{X}}^0)-\tilde{\bm{X}}^*_i  \|_F^2
\end{align}
with learning rate $\alpha_0$.
After that, we further solve problem 
\begin{align}
\mathop{\text{min}}_{\bm{\Theta }_{0:k-1}}
\sum_{i=1}^{T}\| \tilde{\bm{X}}^{k}(\bm{\Theta }_{0:k-1},\tilde{\bm{Y}}_i,
\tilde{\bm{X}}^0)
-\tilde{\bm{X}}^*_i  \|_F^2
\end{align}
to optimize parameters $\bm{\Theta}_{0:k-1}$ with learning rates $\alpha_1$ and $\alpha_2$.
Through the above  process, the first $k$ layers'   parameters can be obtained.
After learning these  parameters, the BS performs JADCE in the test stage by applying the proposed unfolding networks.


\subsubsection{LPGM-AT}
For LPGM-AT,
we only need to find the appropriate hyperparameters (i.e. $c_{\theta}$, $c_{\beta}$, and $c_{\eta}$) by using grid search  in the training stage, which significantly reduces the training cost.

\section{Simulation Results}
In the simulations, the channels between the BS and IoT devices follow independent Rayleigh fading.
The activity of each device follows an independent Bernoulli distribution.
We set  $\mathbb{P}(a_n=0)=0.9$ and $\mathbb{P}(a_n=1)=0.1$, $\forall\  n \in [N]$. 
We set the  regularization parameter $\lambda$ as  $0.1$ and define the transmit SNR as ${\mathbb{E}[\left\| \bm{SX} \right\|^2_F]}/{\mathbb{E}[\left\| \bm{Z} \right\|^2_F]}
$. 
The  neural networks have $K=16$ layers.
The sizes of training dataset, validation dataset, and test dataset are 51200, 2048, and 2048, respectively. 
The learning rates are set to $\alpha_0 = 1\times10^{-3}$, $\alpha_1=0.2\alpha_0$, and $\alpha_2 = 0.02\alpha_0$.
 In the test phase,  the group-sparse-matrix recovery performance is measured  by using  the normalized mean square error (NMSE),  defined as  
 \begin{align}
\text{NMSE}(\tilde{\bm{X}^k},\tilde{\bm{X}}^*)= 10\text{log}_{10}\left(
\frac
{\mathbb{E}\| \tilde{\bm{X}}^k-\tilde{\bm{X}}^* \|^2_F}
{\mathbb{E}\| \tilde{\bm{X}}^* \|^2_F} \right).
\end{align}

\begin{figure}[tbp]
	\centering
	\includegraphics[width=0.55\linewidth]{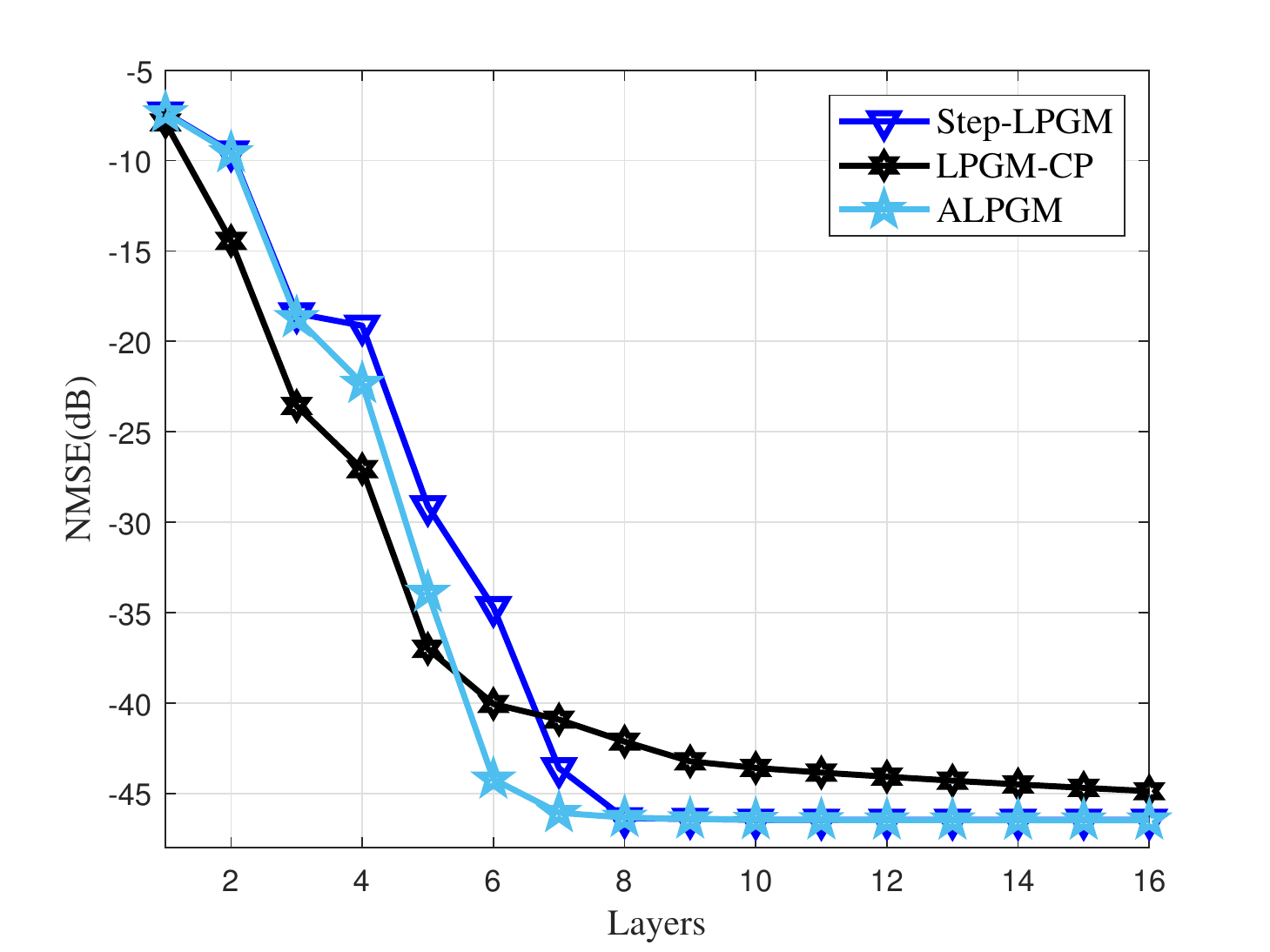}
	\caption{ NMSE versus number of layers for different proximal gradient methods.} 
\end{figure}

\subsection{Performance Comparison}
In the first part of the simulation, ALPGM is compared with the following two unfolding PGM:
\begin{itemize}
	\item Step-LPGM: By fixing  $\tilde{\bm{S}}^{\text{T}}$ in \eqref{eq09} and denoting  $\theta_k = \lambda \gamma_k$, we learn the step-size $\gamma_k$, thresholding parameter $\theta_k$, and parameter $\eta_k$. 
	The trainable parameters are the same as that of ALPGM.  
	The neural network  is given by
	\begin{equation} 
		\begin{aligned}
			\tilde{\bm{X}}^{k+1}\!=\!P_{\theta_k,f_{\eta_k}}\left(\tilde{\bm{X}}^k\! +\! \gamma_k \tilde{\bm{S}}^{\text{T}}(  \tilde{\bm{Y}}\! -\! \tilde{\bm{S}} \tilde{\bm{X}}^k ) \right),\quad k =0,\ldots,K-1.
		\end{aligned}
	\end{equation}
	\item LPGM-CP: We replace $\gamma_k\tilde{\bm{S}}^{\text{T}}$ in \eqref{eq09} by $\bm{B}^k$ and obtain the following neural network 
		\begin{align}
			\tilde{\bm{X}}^{k+1}\!=\!P_{\theta_k,f_{\eta_k}}\left(\tilde{\bm{X}}^k\! +\!  \bm{B}^{k}(  \tilde{\bm{Y}}\! -\! \tilde{\bm{S}} \tilde{\bm{X}}^k ) \right),\quad k =0,\ldots,K-1,
		\end{align}
	where $ \{  \bm{B}^k, \theta_k, \eta_k \} $ are trainable parameters.  
\end{itemize}

We set $M$, $N$, and $L$ to  6, 250, and 125, respectively. 
The SNR is set to 40 dB.
We utilize Zadoff-Chu pilot sequence matrix \cite{chu1972polyphase} and generate it as in \cite{de2022deep}.
Each column of the pilot sequence matrix is normalized.
Fig. 4 shows that our proposed ALPGM achieves a smaller NMSE than LPGM-CP and a  faster convergence rate than Step-LPGM.

\begin{figure*}[tbp]
	\centering
	\subfigure[Complex Gaussian pilot sequence 
	matrix]{
		\includegraphics[width=0.475\linewidth]{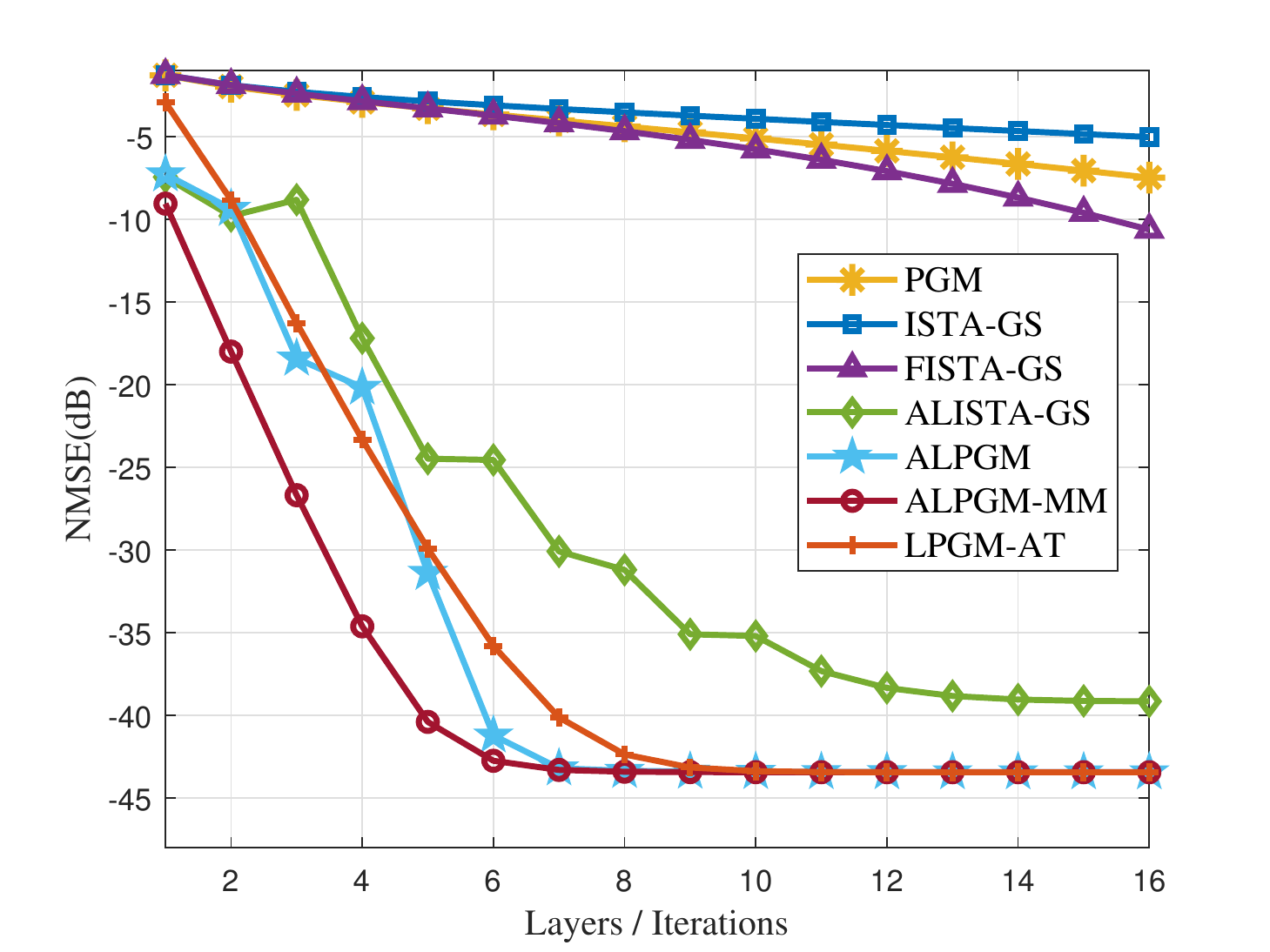}
	}
	\subfigure[Binary pilot sequence 
	matrix]{
		\includegraphics[width=0.475\linewidth]{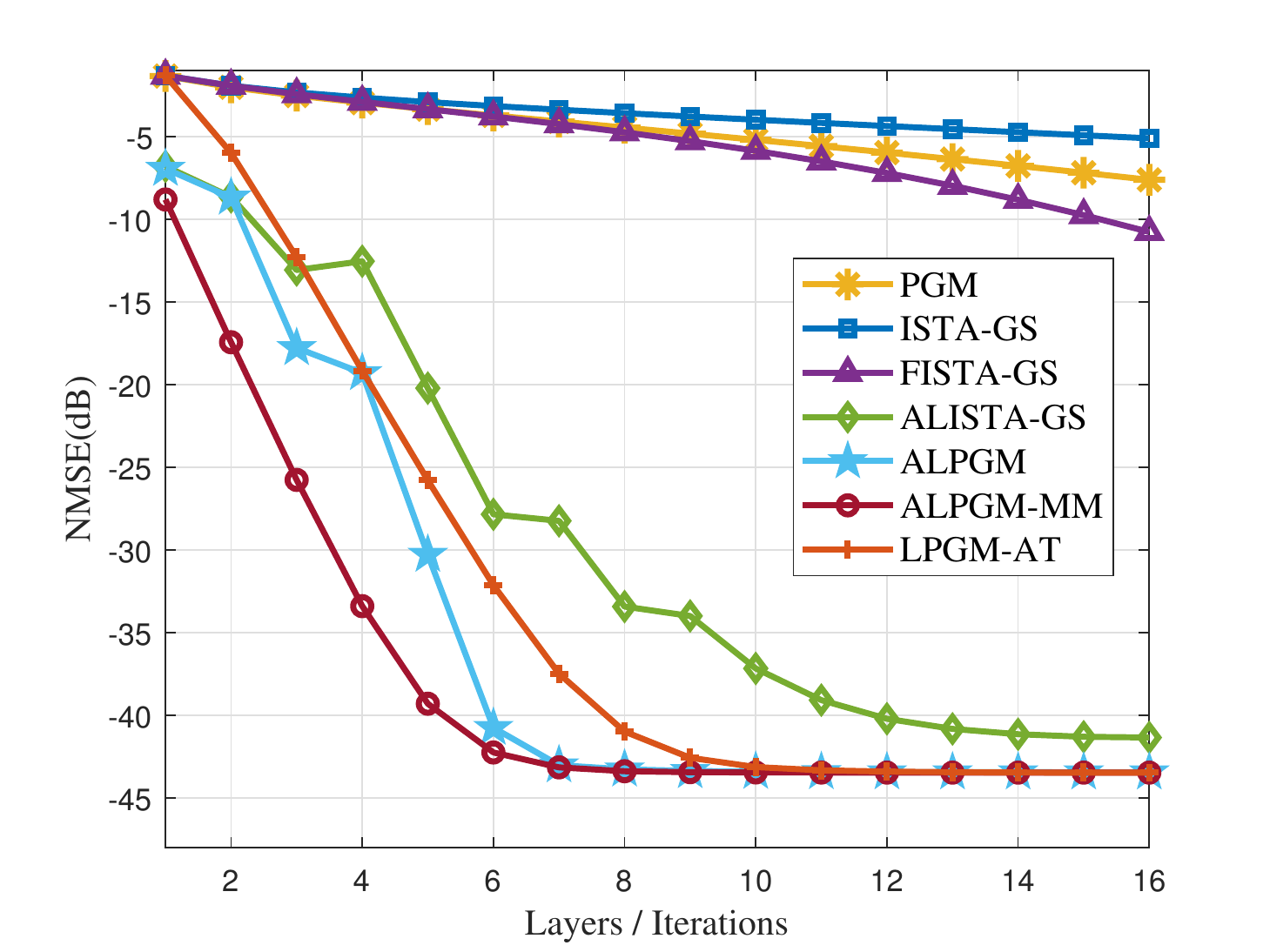}
	}
	\subfigure[Zadoff-Chu pilot
	sequence matrix]{
		\includegraphics[width=0.475\linewidth]{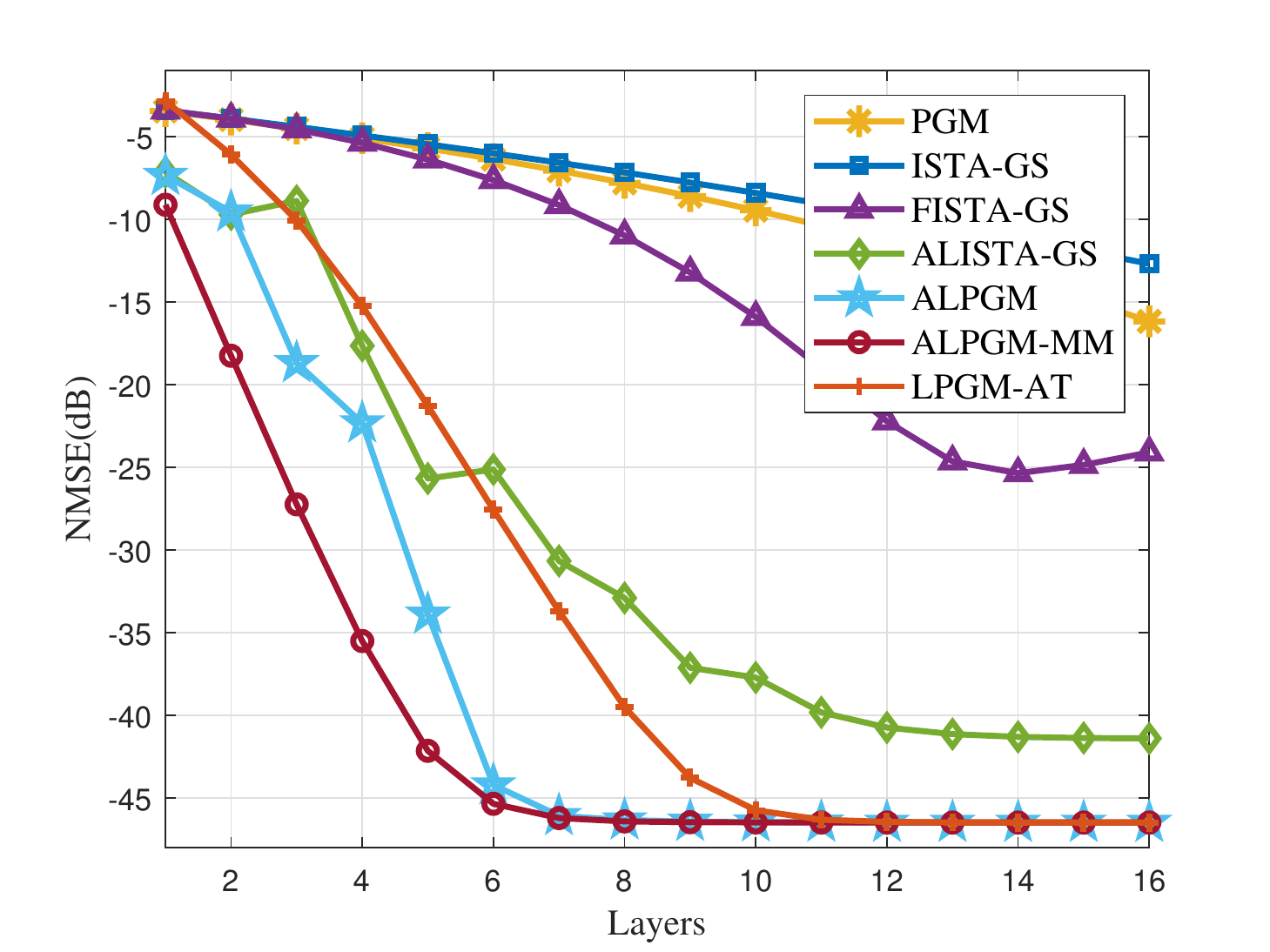}
	}
	\caption{NMSE versus number of layers or iterations for different pilot sequence matrices.}
\end{figure*}

\subsection{Convergence Performance}
	Unfolding PGM has shown its better performance than ISTA and LISTA for solving  SMV problems in \cite{yang2020learning}.
	Thus, we in this paper do not compare the unfolding PGM with these methods for regular sparse recovery. 
	We focus on comparing our proposed structures with the following methods for group sparsity:
\begin{itemize}
	\item \textbf{PGM}: PGM is an iterative algorithm to solve MMV problems.
	The update formula of PGM is given in \eqref{eq09}.
	\item \textbf{ISTA-GS}:
	ISTA-GS \cite{yuan2006model} is an extension of ISTA to solve MMV problems by
	 replacing the scalar soft-thresholding function in ISTA with multidimensional shrinkage thresholding operator.
	The update formula of ISTA-GS is given by
	\begin{align}
		\tilde{\bm{X}}^{k+1}=
		\mathcal{T}_{\lambda/C}\left(\tilde{\bm{X}}^k + \frac{1}{C}\tilde{\bm{S}}^{\text{T}}(\tilde{\bm{Y}} - \tilde{\bm{S}} \tilde{\bm{X}}^k   ) \right),
	\end{align}
	where $\mathcal{T}_{\lambda/C}$ is the multidimensional shrinkage thresholding operator
	\begin{align}
 \mathcal{T}_{\theta}(\tilde{\bm{X}}_{i,:})=\max\{0,
	\|\tilde{\bm{X}}_{i,:}\|_2
	-\theta\}\frac{\tilde{\bm{X}}_{i,:}}{\|\tilde{\bm{X}}_{i,:}\|_2}
\end{align} with  $\theta=\lambda/C$, $\lambda=0.1$, and
	 $C$ denotes the largest eigenvalue of $\tilde{\bm{S}}^{\text{T}}\tilde{\bm{S}}$.
	\item \textbf{Fast ISTA-GS (FISTA-GS)}: 
	FISTA \cite{beck2009fast} is a Nesterov momentum speed-up of  ISTA.
	Correspondingly, FISTA-GS is an accelerated variant  of ISTA-GS to solve MMV problems.
	\item \textbf{ALISTA-GS}: ALISTA-GS is an unfolding algorithm for MMV problems proposed in \cite{shi2021algorithm}.
	The neural network  is 
	\begin{align}
		\tilde{\bm{X}}^{k+1}=\mathcal{T}_{\theta_k}\left(\tilde{\bm{X}}^k + \gamma_k \bm{B} (\tilde{\bm{Y}} - \tilde{\bm{S}} \tilde{\bm{X}}^k   ) \right),
	\end{align}
	where matrix $\bm{B}$ can be obtained by solving problem \eqref{eq16}, and $\{\theta_k,\gamma_k\}$ are trainable parameters.
	Other settings of ALISTA-GS are the same as that of our proposed algorithms. 
\end{itemize}


We evaluate these methods using  three types of pilot sequence matrices, i.e.,  complex Gaussian pilot sequence matrix, binary pilot sequence matrix, and Zadoff-Chu pilot sequence matrix.
Specifically, we generate the complex Guassian pilot sequence matrix by utilizing  the complex Gaussian distribution.
For the binary pilot sequence matrix, each element is selected uniformly at random on $1$ or $-1$.
In addition, each column of the pilot sequence matrix is normalized.
The settings of SNR, device active ratio, $M$, $N$, and $L$ are same as that of Fig. 4.

Fig. 5 depicts the NMSE versus number of layers or iterations for  our proposed networks  and the baseline methods.
ALPGM achieves much lower NMSE than PGM because the parameters in ALPGM are learned to fit the target signals.
Benefiting from the MCP-based multivariate proximal operator, the proposed networks (i.e., ALPGM, ALPGM-MM, and LPGM-AT) achieve   better performance and  faster convergence rate than the baseline methods under all three pilot sequence matrices.
Besides, ALPGM-MM achieves  faster convergence rate than  ALPGM because
the momentum term  accelerates  convergence.
In this experiment, the test data has  the same distribution as the training data, and thus the proposed APGM-AT achieves almost the same performance as ALPGM and ALPGM-MM.

\begin{figure}[tbp] 
	\centering
	\includegraphics[width=0.55\linewidth]{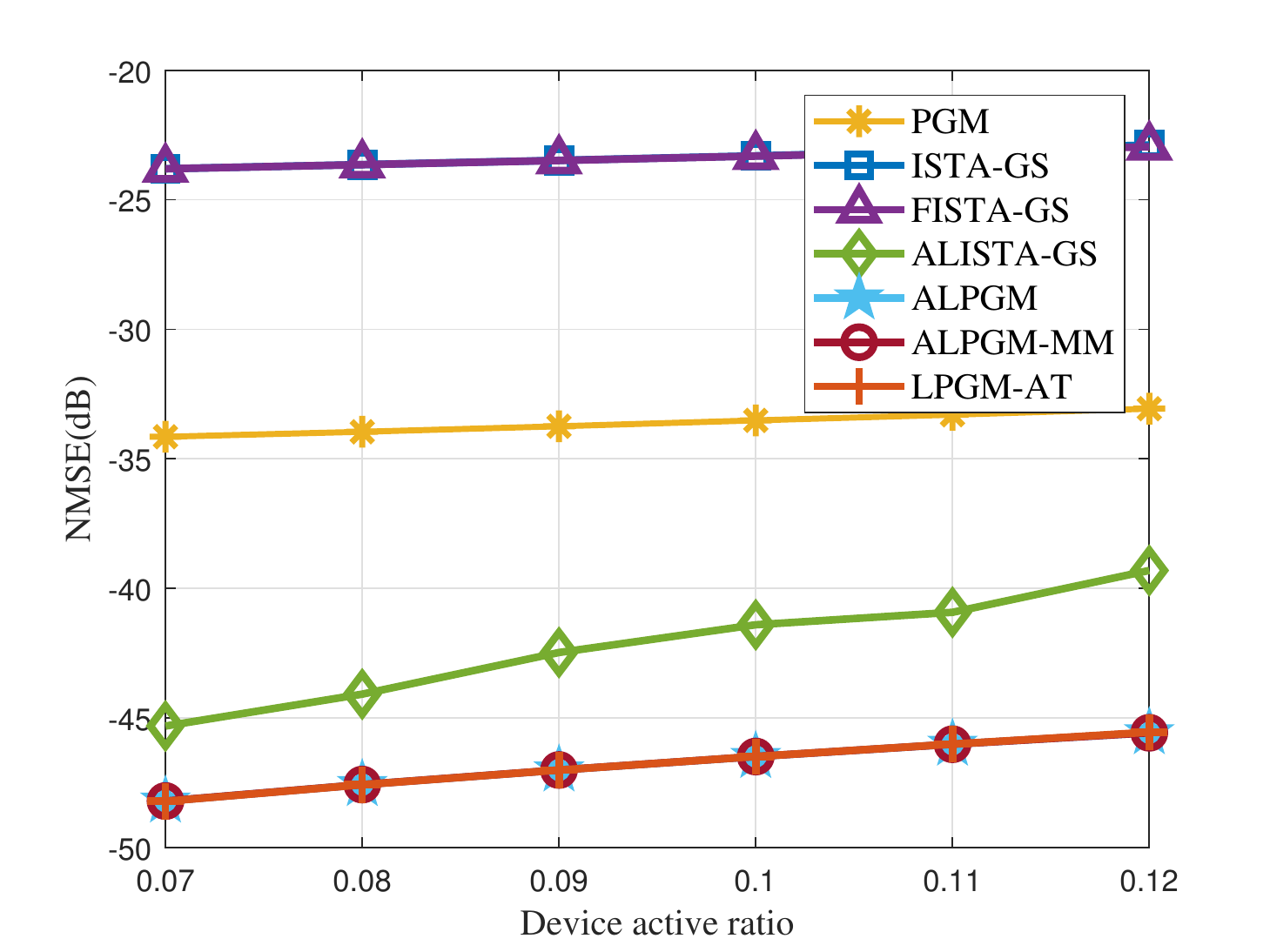}
	\caption{NMSE versus device active ratio when $\bm{S}$ is Zadoff-Chu pilot sequence matrix.} 
\end{figure}


\subsection{Performance Comparison Under Different Settings}
We compare our  proposed three networks with the baselines  under various device active ratios, lengths of pilot, and SNRs.
As the Zadoff-Chu pilot sequence matrix outperforms other practical pilot sequence matrices in Fig. 5, in the following we adopt the Zadoff-Chu pilot sequence matrix.
In order to ensure the convergence of iterative methods (i.e., PGM, ISTA-GS, and FISTA-GS), the numbers  of iterations of iterative methods are set to 50. 
The number of layers of all neural networks is 16.
Other settings are same as Section IV-B.

\begin{figure}[tbp] 
	\centering
	\includegraphics[width=0.55\linewidth]{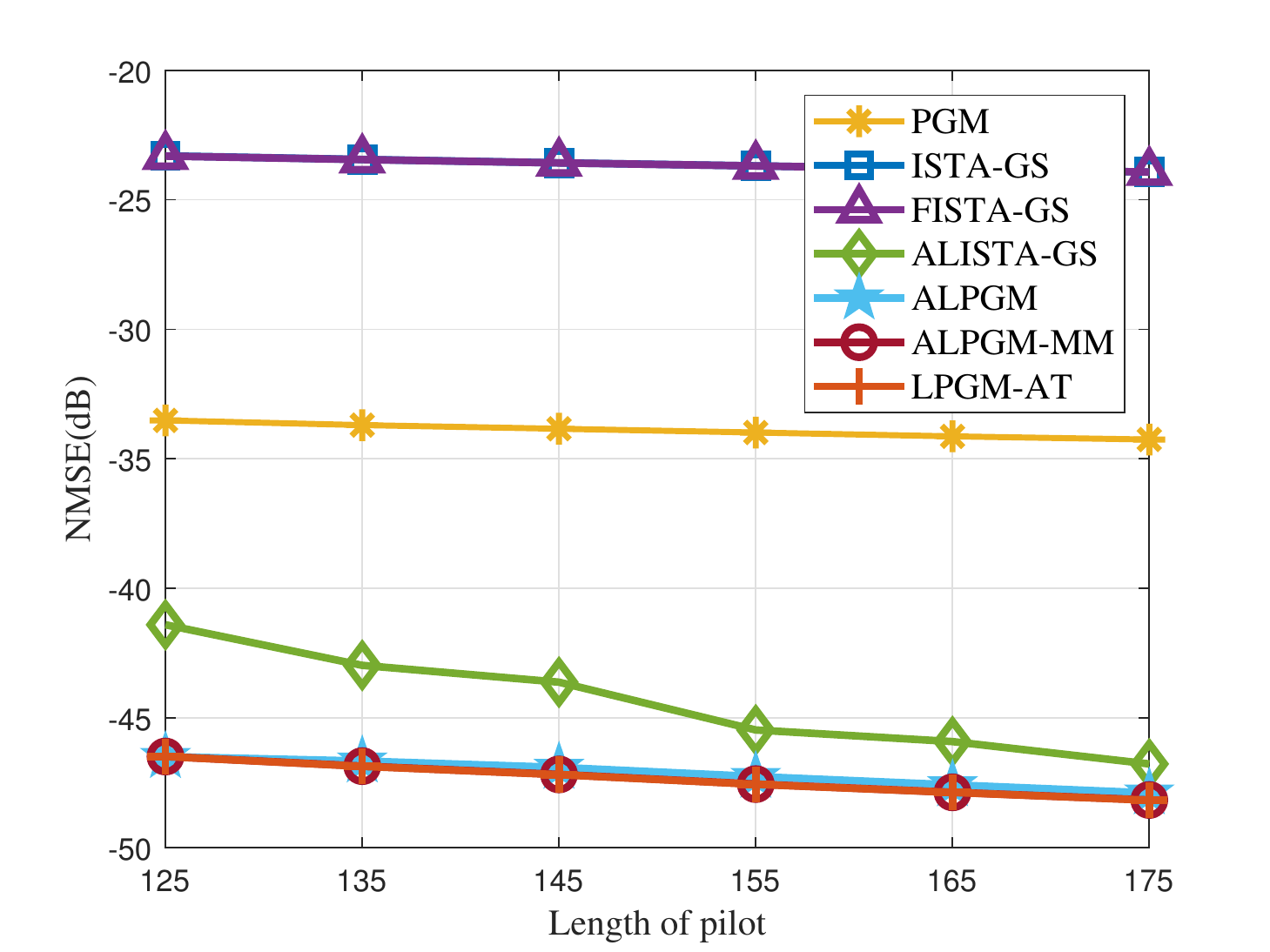}
	\caption{NMSE versus length of pilots when $\bm{S}$ is Zadoff-Chu pilot sequence matrix.} 
\end{figure}

\begin{figure}[tbp] 
	\centering
	\includegraphics[width=0.55\linewidth]{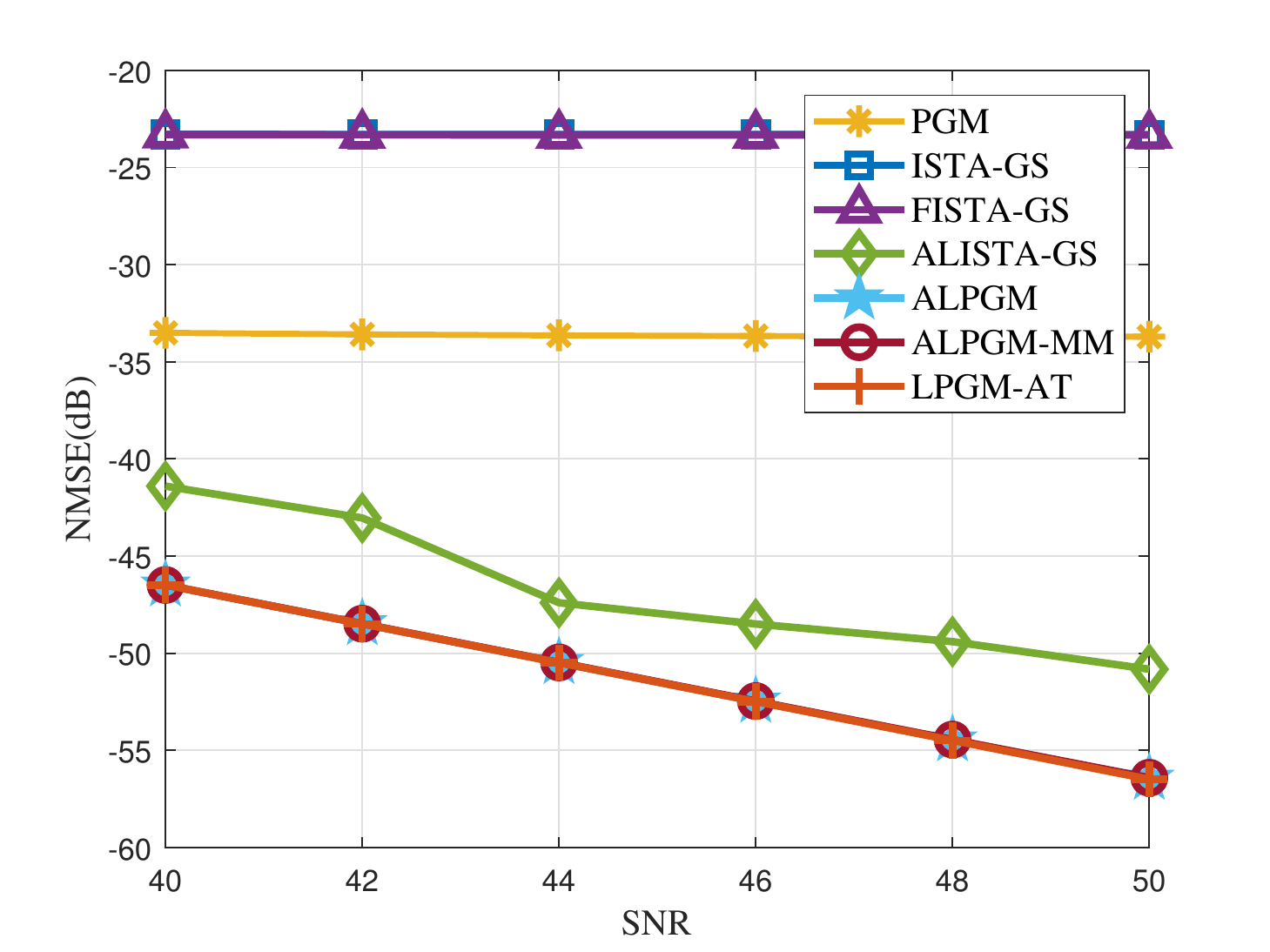}
	\caption{NMSE versus SNR when $\bm{S}$ is Zadoff-Chu pilot sequence matrix.}
\end{figure}

In Figs. 6, 7, and 8, we observe that ISTA-GS and FISTA-GS achieve similar performance after convergence.
In addition, PGM outperforms ISTA-GS because the MCP-based proximal operator is more capable of inducing sparsity than $\ell_1$-norm.
The proposed three networks all achieve much lower NMSEs than the baseline methods under different device active ratios, lengths of pilot, and SNRs.
In Figs. 7 and 8, the NMSE decreases with the length of pilot and SNR, since  
longer pilot sequences and less noise lead to better channel estimation.

The results in Fig. 8 demonstrate that by utilizing MCP, the proposed ALPGM-MM reduces  the NMSE up to $12\%$ compared to  ALISTA-GS when SNR = 40 dB.
Besides, when the test dataset shares the same distribution with the training dataset, the proposed LPGM-AT achieves comparable performance with ALPGM and ALPGM-MM.

\begin{figure*}[tbp]
	\centering
	\subfigure[ SNR of test dataset is changed to 15 dB]{
		\includegraphics[width=0.475\linewidth]{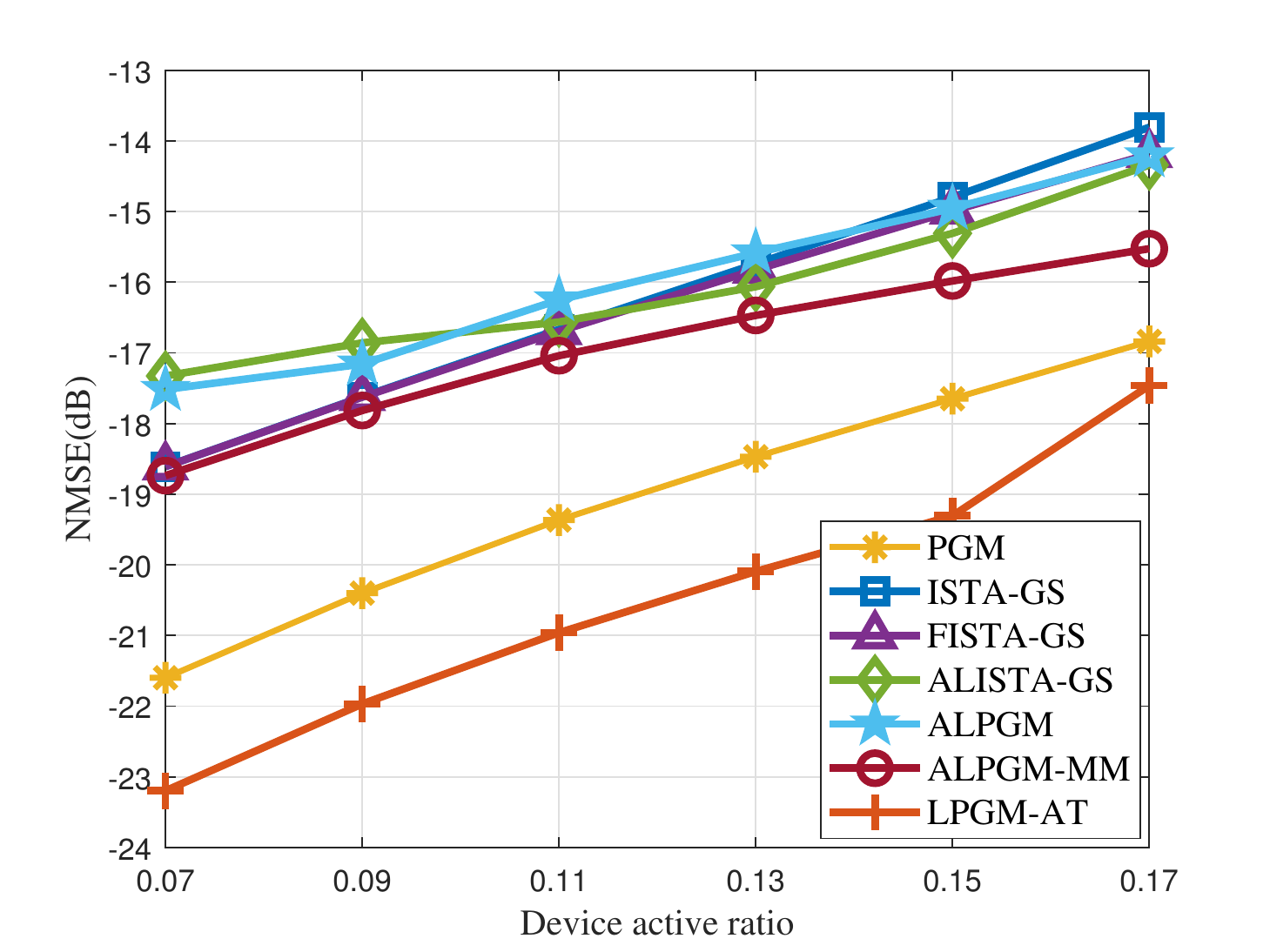}
	}
	\subfigure[ Device active ratio of test dataset is changed to $\mathbb{P}(a_n=1)=0.15$]{
		\includegraphics[width=0.475\linewidth]{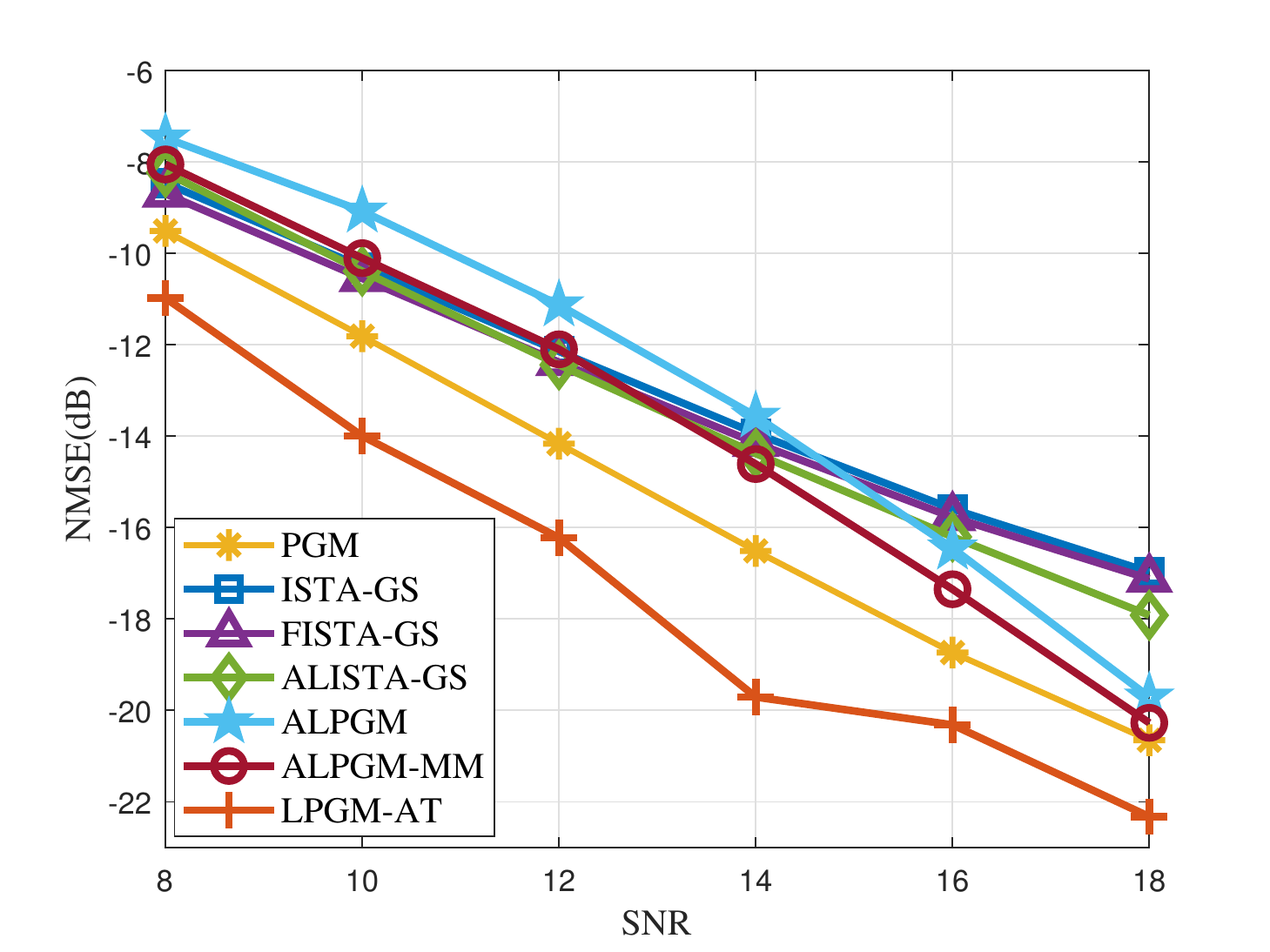}
	}
	\caption{All models except iterative methods are trained when $\bm{S}$ is Zadoff-Chu pilot sequence  matrix and the device active ratio is 0.1 with SNR = 40 dB.}
\end{figure*}

\subsection{Adaptation Comparison}

We compare the adaptivity of the proposed three networks  the baseline methods for the scenario with  mismatch between the training and test datasets.
In this subsection, the  number of iterations of iterative methods (i.e., PGM, ISTA-GS, and FISTA-GS) is set to 50, while the number of the layers of the DL-based methods (i.e., ALPGM, ALPGM-MM, and ALISTA-GS) and LPGM-AT is set to 16.
ALPGM, ALPGM-MM, and ALISTA-GS are trained by back-propagation under the settings of Fig. 5(c).
 LPGM-AT is trained by grid search to find the best hyperparameter combination on the same training dataset as ALPGM, ALPGM-MM, and ALISTA-GS.
Then we directly apply them to the test dataset with different device active ratios and SNRs.

Fig. 9(a) shows the NMSE versus the device active ratio of the test dataset when  SNR $= 15$ dB for  the test dataset. 
The  performance of the DL-based methods (i.e., ALPGM, ALPGM-MM, and ALISTA-GS) degrades  because they are sensitive to the mismatch between the training and test datasets.
However, the results clearly show that LPGM-AT outperforms the DL-based methods and iterative methods because LPGM-AT can adapt its parameters to different distributions of the test dataset.
In Fig. 9(b), we change the device active ratio of the test dataset to 0.15 with lower SNRs than that of the training dataset. 
The results indicate that LPGM-AT is able to adapt to time-varying IoT networks and outperforms  other methods.



\begin{minipage}[htbp]{\textwidth}

	\centering
	\makeatletter\def\@captype{table}\makeatother\caption{Training Time and  Test Time Comparison}
	{
		\resizebox{\linewidth}{!}{
			\begin{tabular}{cccccccc}
				\toprule  
				&ALPGM&ALPGM-MM&LPGM-AT&ALISTA-GS&PGM&ISTA-GS&FISTA-GS\\
				\midrule  
				\tabincell{c}{Training Time} 
				& $2.49$~h& $2.25$~h &$8.18$~min &$3.27$~h&-&-&-\\
				\tabincell{c}{Test Time per Sample} 
				& $6.2\times 10^{-4}$~s
				& $6.5\times 10^{-4}$~s 
				& $6.4\times 10^{-4}$~s 
				& $6.0\times 10^{-4}$~s
				& $2.5\times 10^{-3}$~s
				& $2.3\times 10^{-3}$~s 
				& $2.5\times 10^{-3}$~s\\
				\bottomrule 
			\end{tabular}
		}
	}
	
\end{minipage}

\subsection{Computation Complexity Comparison}

In this subsection,  the training and test time of the proposed three networks are compared  with the baseline methods.
Table I shows that the training time of the DL-based methods (i.e., ALPGM, ALPGM-MM, and ALISTA-GS) need several hours.
This is because  DL-based methods optimize parameters  by back-propagation on a large volume of training data, and the training procedures are time-consuming.
Since the momentum term provides convergence acceleration, ALPGM-MM has the least training time among the DL-based methods.
In contrast, the grid search for LPGM-AT is quite computation-efficient, because it only need to search three hyperparameters.
LPGM-AT only needs less than 10 minutes to find the best hyperparameters by grid search, which dramatically  reduces the computation overhead. 
Moreover, the test time of the proposed network is much smaller than the iterative methods, which demonstrates that the proposed network is more practical for JADCE in IoT networks.

%

\section{Conclusion}
In this paper, we proposed an unfolding framework that is based on PGM for massive random access.
We first mapped PGM as an unfolding neural network to reduce the computational complexity.
In order to further improve the convergence rate, we embedded momentum into
the unfolding neural network, and proved accelerated convergence theoretically.
Based on the convergence analysis, we developed an adaptive network that generalizes  
 well to different device active ratios and SNRs by adjusting its network parameters.
Simulation results showed that the proposed unfolding  framework achieves greater recovery performance, faster convergence, and better adaptivity than the baselines.

\appendix
\subsection{Proof of Theorem 1}

We assume that the noise level $\epsilon$ satisfies
\begin{align}\label{eq41p}
	\epsilon \leq 
	\min\bigg(
	\frac{\underline{\mu_x}}{3\mu_{B}},
	\frac{\underline{\mu_x}(1-c_{\phi s})}{6\mu_{B}(1+s)},
	\frac{\sqrt{s}\mu_x (1-c_{\phi s})(c_{\phi s})^{{K}_0}}{(1+s+\sqrt{s})\mu_{B}}
	\bigg),
\end{align}
where $\hat{\beta}$, $C_0$, and ${K}_0$ are defined as
\begin{align}
	&\hat{\beta}\triangleq\frac{1}{2s}\bigg(1-\sqrt{1-c_{\phi s}}\bigg)^2,
	\\&C_0\triangleq \max\bigg(s \mu_x,
	\frac{8\mu_x s \sqrt{s}(1+\hat{\beta})  }
	{c_{\phi s}\sqrt{4\hat{\beta}-(\hat{\beta}+\phi s -\phi)^2}}
	\bigg),
	\\&{K}_0 \triangleq \bigg\lceil \frac{\log(\underline{\mu_x})-\log(6C_0)}{\log(c_{\phi s})}\bigg\rceil +1.
	\label{eq40p}
\end{align}
Then, we set the specific conditions of parameters  $\{\beta_{k},\eta_{k}\}$ as follows
\begin{align}
	& \beta_k = 
	\left\{
	\begin{aligned}
		&0, &&\text{if}\,k\leq {K}_0,
		\\&\hat{\beta}, &&\text{if}\,k\geq{K}_0+1,
	\end{aligned}
	\right.
	\\&\eta_k 
		\left\{
	\begin{aligned}
		&< \frac{1}{2\theta_k}, &&\text{if}\,k \leq {K}_0 -1,
		\\&= \frac{1}{2\theta_k}, &&\text{if}\,k\geq{K}_0.
	\end{aligned}
	\right.
	\label{eq45p}
\end{align}

\subsubsection{Proof of no-false-positive property} 
When $k=0$ and $\tilde{\bm{X}}^0 = \bm{0}$, we have
$\text{supp}(\psi(\tilde{\bm{X}}^0))=\emptyset \subseteq S$.
We fix $k\geq0$ and assume $\bm{X}^t_{i,:}=0,\forall\,i\notin S, 0\leq t \leq k$.
For $\forall\, i\notin S$, we have
\begin{align} \label{eq46p}
	&\| -  \bm{B}_{i,:}\tilde{\bm{S}}_{:,S}(\tilde{\bm{X}}^k_{S,:} - \tilde{\bm{X}}^*_{S,:}) \|_2 
	= 
	\|-\sum_{l\in S}\bm{B}_{i,:}\tilde{\bm{S}}_{:,l}(\tilde{\bm{X}}^k_{l,:} - \tilde{\bm{X}}^*_{l,:}) \|_2 \notag
	\\&\leq
	\sum_{l\in S} \|\bm{B}_{i,:}\tilde{\bm{S}}_{:,l}(\tilde{\bm{X}}^k_{l,:} - \tilde{\bm{X}}^*_{l,:}) \|_2 
	\leq
	\sum_{l\in S} |\bm{B}_{i,:}\tilde{\bm{S}}_{:,l}|
	\|\tilde{\bm{X}}^k_{l,:} - \tilde{\bm{X}}^*_{l,:} \|_2 \notag
	\\&\mathop{\leq}^{(a)}
	\sum_{l\in S} \phi
	\|\tilde{\bm{X}}^k_{l,:} - \tilde{\bm{X}}^*_{l,:} \|_2 
	\leq
	\phi \|\tilde{\bm{X}}^k - \tilde{\bm{X}}^* \|_{2,1},
\end{align}
where (a) is due to  $
\phi \geq  |(\bm{D}^\text{T}\bm{D})_{i,l}|=|\bm{B}_{i,:}\tilde{\bm{S}}_{:,l}|
$.
Since $\|\bm{B}\|_{F}\leq\|\bm{B}\|_{2,1}\leq \mu_{B}$ and $ \| \tilde{\bm{Z}} \|_F \leq \epsilon$, we have
\begin{align} \label{eq47p}
	\|\bm{B}_{i,:} \tilde{\bm{Z}}\|_2
	\leq \|\bm{B}_{i,:}\|_{2}\| \tilde{\bm{Z}} \|_F
	\leq \|\bm{B}\|_{F}\| \tilde{\bm{Z}} \|_F
	\leq \mu_{B} \epsilon.
\end{align}
By combining \eqref{eq46p} and \eqref{eq47p}, we obtain the lower bound for the threshold parameter $\theta_k$
\begin{align}\label{eq48p}
	\frac{1}{2\eta_k} >
	\theta_k
	&\geq \phi\| \tilde{\bm{X}}^k - \tilde{\bm{X}}^* \|_{2,1} + \mu_{B}\epsilon \notag
	\\&\geq  
	\|-\bm{B}_{i,:}\tilde{\bm{S}}_{:,S}(\tilde{\bm{X}}^k_{S,:} - \tilde{\bm{X}}^*_{S,:}) \|_2
	+ \| \bm{B}_{i,:} \tilde{\bm{Z}} \|_2 \notag
	\\&\geq  \|-  \bm{B}_{i,:}\tilde{\bm{S}}_{:,S}(\tilde{\bm{X}}^k_{S,:} - \tilde{\bm{X}}^*_{S,:}) + \bm{B}_{i,:} \tilde{\bm{Z}} \|_2.
\end{align}

From the update formula of ALPGM-MM, for $\forall \, i\notin S$, we have
\begin{align}
	\tilde{\bm{X}}_{i,:}^{k+1}
	&=P_{\theta_k,f_{\eta_k}}\bigg(\tilde{\bm{X}}_{i,:}^k -  \bm{B}_{i,:}(    \tilde{\bm{S}} \tilde{\bm{X}}^k - \tilde{\bm{Y}})
	+ \beta^k(\tilde{\bm{X}}_{i,:}^k-\tilde{\bm{X}}_{i,:}^{k-1}) \bigg)\notag
	\\&
	=P_{\theta_k,f_{\eta_k}}\bigg(\tilde{\bm{X}}_{i,:}^k -  \bm{B}_{i,:}\tilde{\bm{S}}_{:,S}(      \tilde{\bm{X}}_{S,:}^k - \tilde{\bm{X}}_{S,:}^*)
	+\bm{B}_{i,:}\tilde{\bm{Z}}
	+ \beta_k(\tilde{\bm{X}}_{i,:}^k-\tilde{\bm{X}}_{i,:}^{k-1}) \bigg) .
\end{align}
As $\bm{X}^t_{i,:}=0,\forall\,i\notin S, 0\leq t \leq k$, we obtain
\begin{align}
	\tilde{\bm{X}}_{i,:}^{k+1}
	=P_{\theta_k,f_{\eta_k}}\bigg( -  \bm{B}_{i,:}\tilde{\bm{S}}_{:,S}(\tilde{\bm{X}}^k_{S,:} - \tilde{\bm{X}}^*_{S,:}) + \bm{B}_{i,:} \tilde{\bm{Z}} \bigg)
	= \hat{P}_{\theta_k,f_{\eta_k}}(\|\bm{v}\|_2)\frac{\bm{v}}{\|\bm{v}\|_2},
\end{align}
where $ \bm{v} = -  \bm{B}_{i,:}\tilde{\bm{S}}_{:,S}(\tilde{\bm{X}}^k_{S,:} - \bm{X}^*_{S,:}) + \bm{B}_{i,:} \tilde{\bm{Z}}$.
According to \eqref{eq48p} and \eqref{eq12}, we obtain $\tilde{\bm{X}}^{k+1}_{i,:}=0, \forall\, i \notin  S$.
By induction, we complete the proof.

\subsubsection{Convergence analysis}
{Firstly, we analyze the convergence when $\beta_k=0$.}
When $\beta_k=0$ and $k\leq {K}_0$, 
ALPGM-MM reduces to ALPGM. 
By the definition of multivariate proximal operator, for $\forall\,i\in S$, we have
\begin{align}
	&\tilde{\bm{X}}^{k+1}_{i,:} 
	= P_{\theta_k,f_{\eta_k}}\left(
	\tilde{\bm{X}}^{k}_{i,:}  - 
	\bm{B}_{i,:}\tilde{\bm{S}}_{:,S} 
	(\tilde{\bm{X}}^k_{S,:}-\tilde{\bm{X}}^{*}_{S,:}) 
	+  \bm{B}_{i,:}\tilde{\bm{Z}}
	+\beta_k(\tilde{\bm{X}}_{i,:}^k-\tilde{\bm{X}}_{i,:}^{k-1})
	\right) \notag
	\\& =\! \arg\min_{\tilde{\bm{U}}_{i,:}} 
	\frac{1}{2} \| \tilde{\bm{U}}_{i,:} 
	\!-\! (\tilde{\bm{X}}^{k}_{i,:} \!-\! 
	\bm{B}_{i,:}\tilde{\bm{S}}_{:,S} 
	(\tilde{\bm{X}}^k_{S,:}\!-\!\tilde{\bm{X}}^{*}_{S,:}) 
	\!+\!  \bm{B}_{i,:}\tilde{\bm{Z}}
	\!+\! \beta_k(\tilde{\bm{X}}_{i,:}^k-\tilde{\bm{X}}_{i,:}^{k-1})
	) \|_2^2 
	+ \theta_k  g_{\eta_k}(\|\tilde{\bm{U}}_{i,:}\|_2).
\end{align}
According to the optimality condition, we have
\begin{align}\label{eq51p}
	\bm{0} \in  \tilde{\bm{X}}^{k+1}_{i,:} 
	- \left(\tilde{\bm{X}}^{k}_{i,:}  - 
	\bm{B}_{i,:}\tilde{\bm{S}}_{:,S} 
	(\tilde{\bm{X}}^k_{S,:}-\tilde{\bm{X}}^{*}_{S,:}) 
	+  \bm{B}_{i,:}\tilde{\bm{Z}}
	+ \beta_k(\tilde{\bm{X}}_{i,:}^k-\tilde{\bm{X}}_{i,:}^{k-1})
	\right)
	+ \theta_k \partial g_{\eta_k}(\|\tilde{\bm{X}}_{i,:}^{k+1}\|_2),
\end{align}
where $\partial g_{\eta_k}(\|\tilde{\bm{X}}_{i,:}^{k+1}\|_2)$ is the subgradient of $g_{\eta_k}(\|\tilde{\bm{X}}_{i,:}^{k+1}\|_2)$.

Recalling the definition of $g_{\eta}(\cdot)$, we have
\begin{align}
	g_{\eta_k}(\|\tilde{\bm{X}}_{i,:}^{k+1}\|_2)
	= \bigg(\|\tilde{\bm{X}}_{i,:}^{k+1}\|_2 - \eta_k \|\tilde{\bm{X}}_{i,:}^{k+1}\|_2^2\bigg)
	\mathbbm{1}_{\|\tilde{\bm{X}}_{i,:}^{k+1}\|_2\leq \frac{1}{2\eta_k}}(\|\tilde{\bm{X}}_{i,:}^{k+1}\|_2) \notag
	\\ + \bigg(\frac{1}{4\eta_k}\bigg)
	\mathbbm{1}_{\|\tilde{\bm{X}}_{i,:}^{k+1}\|_2 > \frac{1}{2\eta_k}}(\|\tilde{\bm{X}}_{i,:}^{k+1}\|_2).
\end{align}
One can easily check that
\begin{align}
	\partial g_{\eta_k}(\|\tilde{\bm{X}}_{i,:}^{k+1}\|_2)
	=  \bigg( \partial\|\tilde{\bm{X}}_{i,:}^{k+1}\|_2 
	- 2\eta_k \tilde{\bm{X}}_{i,:}^{k+1} \bigg)
	\mathbbm{1}_{\|\tilde{\bm{X}}_{i,:}^{k+1}\|_2\leq \frac{1}{2\eta_k}},
\end{align}
where 
\begin{align}
	\partial \| \tilde{\bm{X}}_{i,:}^{k+1}\|_2
	= \left\{\begin{aligned}
	&	\frac{\tilde{\bm{X}}_{i,:}^{k+1}}{\|\tilde{\bm{X}}_{i,:}^{k+1}\|_2},
	&&\text{if}\,\tilde{\bm{X}}_{i,:}^{k+1}\neq \bm{0},
	\\& \{\bm{h}\in\mathbb{R}^{1\times M}|\|\bm{h}\|_2\leq 1\},
	&& \text{otherwise}.
	\end{aligned}\right.
\end{align}
Hence, we obtain
\begin{align}
	&\| \partial g_{\eta_k}(\|\tilde{\bm{X}}_{i,:}^{k+1}\|_2) \|_2^2 \notag
	\\&=\bigg(\|\bm{h}\|^2_2\bigg)\mathbbm{1}_{\|\tilde{\bm{X}}_{i,:}^{k+1}\|_2=0}(\|\tilde{\bm{X}}_{i,:}^{k+1}\|_2)
	+\bigg(\bigg\|\frac{\tilde{\bm{X}}_{i,:}^{k+1}}{\|\tilde{\bm{X}}_{i,:}^{k+1}\|_2}-2\eta_k\tilde{\bm{X}}_{i,:}^{k+1}\bigg\|^2_2\bigg)
	\mathbbm{1}_{0<\|\tilde{\bm{X}}_{i,:}^{k+1}\|_2\leq
	\frac{1}{2\eta_k}}(\|\tilde{\bm{X}}_{i,:}^{k+1}\|_2)\notag
	\\&\leq \mathbbm{1}_{\|\tilde{\bm{X}}_{i,:}^{k+1}\|_2=0}(\|\tilde{\bm{X}}_{i,:}^{k+1}\|_2)
	+
	\mathbbm{1}_{0<\|\tilde{\bm{X}}_{i,:}^{k+1}\|_2\leq
		\frac{1}{2\eta_k}}(\|\tilde{\bm{X}}_{i,:}^{k+1}\|_2)
	=1.
\end{align}

From \eqref{eq51p}, we have
\begin{align}\label{eq56p}
	\tilde{\bm{X}}^{k+1}_{i,:} -  \tilde{\bm{X}}^{*}_{i,:}
	&= \tilde{\bm{X}}^{k}_{i,:}  -  \tilde{\bm{X}}^{*}_{i,:}
	- \bm{B}_{i,:}\tilde{\bm{S}}_{:,S} 
	(\tilde{\bm{X}}^k_{S,:}-\tilde{\bm{X}}^{*}_{S,:}) 
	+  \bm{B}_{i,:}\tilde{\bm{Z}}
	-  \theta_k \partial g_{\eta_k}(\|\tilde{\bm{X}}_{i,:}^{k+1}\|_2) \notag
	\\&= 
	- \sum_{j\in S,\,j\neq i} \bm{B}_{i,:}\tilde{\bm{S}}_{:,j} 
	(\tilde{\bm{X}}^k_{j,:}-\tilde{\bm{X}}^{*}_{j,:}) 
	+  \bm{B}_{i,:}\tilde{\bm{Z}}
	-  \theta_k \partial g_{\eta_k}(\|\tilde{\bm{X}}_{i,:}^{k+1}\|_2),
\end{align}
where the last equality follows from  the constraint $(\bm{D}^{\text{T}}\bm{D})_{i,i}=\bm{B}_{i,:}\tilde{\bm{S}}_{:,i}=1$.

We take  norm on both sides of \eqref{eq56p} and obtain
\begin{align}
	\|\tilde{\bm{X}}^{k+1}_{i,:} -  \tilde{\bm{X}}^{*}_{i,:}\|_2
	&\leq
	\sum_{j\in S,\,j\neq i} | \bm{B}_{i,:}\tilde{\bm{S}}_{:,j} |
	\| \tilde{\bm{X}}^k_{j,:}-\tilde{\bm{X}}^{*}_{j,:} \|_2 
	+  \| \bm{B}_{i,:}\tilde{\bm{Z}} \|_2
	+  \theta_k \| \partial g_{\eta_k}(\|\tilde{\bm{X}}_{i,:}^{k+1}\|_2) \|_2 \notag
	\\&\leq
	\phi \sum_{j\in S,\,j\neq i} 
	\| \tilde{\bm{X}}^k_{j,:}-\tilde{\bm{X}}^{*}_{j,:} \|_2 
	+  \| \bm{B}_{i,:}\tilde{\bm{Z}} \|_2
	+  \theta_k .
\end{align}
Due to the no-false-positive property, we obtain $\|\tilde{\bm{X}}^{k+1} - \tilde{\bm{X}}^{*}\|_{2,1} = \|\tilde{\bm{X}}^{k+1}_{S,:} - \tilde{\bm{X}}^{*}_{S,:}\|_{2,1}$ and 
\begin{align}\label{eq58p}
	\|\tilde{\bm{X}}^{k+1} - \tilde{\bm{X}}^{*}\|_{2,1}
	\leq 
	\phi(|S|-1) \| \tilde{\bm{X}}^k-\tilde{\bm{X}}^{*} \|_{2,1}
	 + \mu_{B} \epsilon
	 + |S|\theta_{k}.
\end{align}
By taking supremum on both sides of inequality \eqref{eq58p}, we obtain
\begin{align}\label{eq59p}
	\mathop{\sup}_{(\tilde{\bm{X}}^*,\tilde{\bm{Z}})\in\mathcal{X}(\underline{\mu_x},\mu_x,s,\epsilon)}
	\|\tilde{\bm{X}}^{k+1} - \tilde{\bm{X}}^{*}\|_{2,1}
	\leq 
	\phi(s-1)\mathop{\sup}_{(\tilde{\bm{X}}^*,\tilde{\bm{Z}})\in\mathcal{X}(\underline{\mu_x},\mu_x,s,\epsilon)}
	 \| \tilde{\bm{X}}^k-\tilde{\bm{X}}^{*} \|_{2,1}
	+ \mu_{B} \epsilon
	+ s\theta_{k}.
\end{align}                                             
By plugging the definition of $\theta_k$ into \eqref{eq59p}, 
we obtain
\begin{align}
&\mathop{\sup}_{(\tilde{\bm{X}}^*,\tilde{\bm{Z}})\in\mathcal{X}(\underline{\mu_x},\mu_x,s,\epsilon)}
\|\tilde{\bm{X}}^{k+1} - \tilde{\bm{X}}^{*}\|_{2,1}\notag
\\&\leq\!
\phi(s-1)\mathop{\sup}_{(\tilde{\bm{X}}^*,\tilde{\bm{Z}})\in\mathcal{X}(\underline{\mu_x},\mu_x,s,\epsilon)}
\| \tilde{\bm{X}}^k-\tilde{\bm{X}}^{*} \|_{2,1}
\!+\! \mu_{B} \epsilon
\!+\!\phi s  \mathop{\text{sup}}_{(\tilde{\bm{X}^*},\tilde{\bm{Z}})\in\mathcal{X}(\underline{\mu_x},\mu_x,s,\epsilon)  }  \| \tilde{\bm{X}}^k  - \tilde{ \bm{X}}^* \|_{2,1} 
+ s \mu_{B} \epsilon  \notag
\\&=
(2\phi s-\phi)\mathop{\sup}_{(\tilde{\bm{X}}^*,\tilde{\bm{Z}})\in\mathcal{X}(\underline{\mu_x},\mu_x,s,\epsilon)}
\| \tilde{\bm{X}}^k-\tilde{\bm{X}}^{*} \|_{2,1}
+ (1+s)\mu_{B} \epsilon.
\end{align}
We denote $e^k=\mathop{\text{sup}}_{(\tilde{\bm{X}}^*,\tilde{\bm{Z}})\in\mathcal{X}(\underline{\mu_x},\mu_x,s,\epsilon)  }  \| \tilde{\bm{X}}^k  - \tilde{ \bm{X}}^* \|_{2,1}$.
If $(2s-1)\phi < 1$,
we obtain
\begin{align}
	e^{k+1} 
	&\leq 
	(c_{\phi s}) e^{k} 
	+ (1+s) \mu_{B} \epsilon \notag
	\\&\leq
	 (c_{\phi s})^{k+1} e^0 
	+ \bigg(\sum_{t=0}^{k}(c_{\phi s})^t\bigg)
	(1+s) \mu_{B} \epsilon  \notag 
	\\&\leq
	 (c_{\phi s})^{k+1} e^0 
	+  \frac{(1+s) \mu_{B} \epsilon}{1-c_{\phi s}} .
\end{align}
As $\tilde{\bm{X}}^0=0$, we obtain
\begin{align}
	e^0 = \mathop{\sup}_{(\tilde{\bm{X}}^*,\tilde{\bm{Z}})\in\mathcal{X}(\underline{\mu_x},\mu_x,s,\epsilon)}
	\|\tilde{\bm{X}}^{*}\|_{2,1}
	\leq s\mu_x 
	\leq C_0.
\end{align}
Since $\|\bm{X}\|_F\leq\|\bm{X}\|_{2,1}$, we conclude with 
\begin{align}\label{eq60p}
	\| \tilde{\bm{X}}^{k}  - \tilde{ \bm{X}}^* \|_{F}
	\leq C_0 (c_{\phi s})^{k}
	+ \frac{(1+s) \mu_{B} \epsilon}{1-c_{\phi s}}, 
	\quad 
	\forall\, k \leq {K}_0+1.
\end{align}

{Secondly, we analyze the convergence when $\beta_k=\hat{\beta}$.}
We define $\bar{\bm{X}}_{S,:} = \tilde{\bm{X}}_{S,:}^*+\delta\tilde{\bm{X}}_{S,:}$, where $\delta\tilde{\bm{X}}_{S,:} = (\bm{B}_{S,:}\tilde{\bm{S}}_{:,S})^{-1}\bm{B}_{S,:}\tilde{\bm{Z}}$.
Then, we obtain
\begin{align}
	\|(\bm{B}_{S,:}\tilde{\bm{S}}_{:,S})^{-1}\|_F
	\mathop{\leq}^{(a)} \sqrt{|S|}\|(\bm{B}_{S,:}\tilde{\bm{S}}_{:,S})^{-1}\|_2
	= \frac{\sqrt{|S|}}{\sigma_{\min}(\bm{B}_{S,:}\tilde{\bm{S}}_{:,S})}
	\mathop{\leq}^{(b)} \frac{\sqrt{|S|}}{1+\phi-\phi|S|}
	\leq \frac{\sqrt{s}}{1+\phi-\phi s},
\end{align}
where (a) is due to $
\|\bm{A}\|_F^2 = \sum_{i=1}^{n}\sigma_i^2(\bm{A}) 
\leq n  \sigma^{2}_{\max}(\bm{A})
= n \|\bm{A}\|_2^2
$
with $\sigma_i(\bm{A})$ denoting the  singular value of the  matrix $\bm{A}\in\mathbb{R}^{n\times n}$,
and (b) follows by 
Gershgorin circle theorem [Chapter 7]\cite{golub2013matrix}.
Hence, we obtain
\begin{align}\label{eq65p}
	\| \delta\tilde{\bm{X}}_{S,:} \|_F
	\leq \|(\bm{B}_{S,:}\tilde{\bm{S}}_{:,S})^{-1}\|_F
	\|\bm{B}_{S,:}\tilde{\bm{Z}}\|_F
	\leq 
	\frac{\sqrt{s} \mu_{B}\epsilon}{1+\phi-\phi s}
	.
\end{align}

From \eqref{eq51p}, for $k \geq \tilde{K_0}+1$, we have
\begin{align}\label{eq63p}
	\tilde{\bm{X}}^{k+1}_{S,:} = \tilde{\bm{X}}^{k}_{S,:}  - 
	\bm{B}_{S,:}\tilde{\bm{S}}_{:,S} 
	(\tilde{\bm{X}}^k_{S,:}-\bar{\bm{X}}_{S,:}) 
	+ \beta_k (\tilde{\bm{X}}^k_{S,:}-\tilde{\bm{X}}^{k-1}_{S,:})
 - \theta_k \partial g_{\eta_k}(\|\tilde{\bm{X}}_{S,:}^{k+1}\|_2),
\end{align}
where $\partial g_{\eta_k}(\|\tilde{\bm{X}}_{2N,:}^{k+1}\|_2)$ is defined as 
\begin{align}
	\partial g_{\eta_k}(\|\tilde{\bm{X}}_{2N,:}^{k+1}\|_2) = \bigg[
	\partial g_{\eta_k}(\|\tilde{\bm{X}}_{1,:}^{k+1}\|_2)^{\text{T}},
	\ldots,
	\partial g_{\eta_k}(\|\tilde{\bm{X}}_{2N,:}^{k+1}\|_2)^{\text{T}}
	\bigg]^{\text{T}}.
\end{align}
By substracting $\bar{\bm{X}}_{S,:}$ from both sides of \eqref{eq63p}, we obtain
\begin{align}\label{eq64p}
	\tilde{\bm{X}}^{k+1}_{S,:}-\bar{\bm{X}}_{S,:}
	\!=\! \bigg((1+\beta_k)\bm{I}_S - \bm{B}_{S,:}\tilde{\bm{S}}_{:,S}\bigg)(\tilde{\bm{X}}^k_{S,:}-\bar{\bm{X}}_{S,:})
	\!-\!\beta_k (\tilde{\bm{X}}^{k-1}_{S,:}\!-\!\bar{\bm{X}}_{S,:}) 
	- \theta_k \partial g_{\eta_k}(\|\tilde{\bm{X}}_{S,:}^{k+1}\|_2).
\end{align}
By plugging $\beta_k=\hat{\beta}$ with $k \geq {K}_0+1$ into \eqref{eq64p},
we obtain
\begin{align} \label{eq68p}
	\underbrace{
	\begin{bmatrix}
		\tilde{\bm{X}}^{k+1}_{S,:}-\bar{\bm{X}}_{S,:}
		\\	\tilde{\bm{X}}^{k}_{S,:}-\bar{\bm{X}}_{S,:}
	\end{bmatrix}
	}_{\bm{z}^k}
	= 
	\underbrace{
	\begin{bmatrix}
		&(1+\hat{\beta})\bm{I}_S - \bm{B}_{S,:}\tilde{\bm{S}}_{:,S}
		&& -\hat{\beta}\bm{I}_S
		\\&\bm{I}_S
		&&\bm{0}
	\end{bmatrix}
	}_{\bm{M}}
	\underbrace{
	\begin{bmatrix}
		\tilde{\bm{X}}^{k}_{S,:}-\bar{\bm{X}}_{S,:}
		\\	\tilde{\bm{X}}^{k-1}_{S,:}-\bar{\bm{X}}_{S,:}
	\end{bmatrix}
	}_{\bm{z}^{k-1}}
	-
	\theta_{k}
	\begin{bmatrix}
		\partial g_{\eta_k}(\|\tilde{\bm{X}}_{S,:}^{k+1}\|_2)
		\\ \bm{0}
	\end{bmatrix}.
\end{align}
There exist a nonsingular matrix $\bm{T}\in\mathbb{C}^{2|S|\times2|S|}$ and a diagonal matrix $\bm{\Lambda_{\bm{M}}}\in\mathbb{C}^{2|S|\times2|S|}$ such that matrix $\bm{M}$ can be factorized as
$\bm{M} = \bm{T}\bm{\Lambda_{\bm{M}}}\bm{T}^{-1}$.
The matrices $\bm{T}$ and $\bm{\Lambda_{\bm{M}}}$ satisfy
\begin{align}
	&\|\bm{\Lambda_{\bm{M}}}\|_F= \sqrt{2|S|\hat{\beta}}
	\leq \sqrt{2s\hat{\beta}},
	\\& \|\bm{T} \|_F=\sqrt{|S|(2+2\hat{\beta})}
	\leq \sqrt{2s(1+\hat{\beta})},
	\\& \|\bm{T}^{-1}\|_F 
	\leq \sqrt{\frac{|S|(2+2\hat{\beta})}{4\hat{\beta}-(\hat{\beta}+\phi s -\phi)^2}}
	\leq
	\sqrt{\frac{2s(1+\hat{\beta})}{4\hat{\beta}-(\hat{\beta}+\phi s -\phi)^2}}.
\end{align}
The proof follows the idea in \cite{chen2021hyperparameter} with some modifications according to the problem that we consider.

%

Before we estimate the recovery error $\|\tilde{\bm{X}}^{k+1}-\tilde{\bm{X}}^*\|_F$, 
we use induction to prove  $\partial g_{\eta_t}(\|\tilde{\bm{X}}_{S,:}^{t+1}\|_2)\\=\bm{0}$  for all $t$ satisfying ${K}_0 \leq t \leq k \,(k\geq{K}_0)$.

(i) We prove  that $\partial g_{\eta_{{{K}_0}}}(\|\tilde{\bm{X}}_{S,:}^{{{K}_0+1}}\|_2)=\bm{0}$.
According to the definition of ${K}_0$ \eqref{eq40p}, 
 it holds that $C_0 (c_{\phi s})^{k} < \underline{\mu_x}/6$ for  $\forall \, k\geq{K}_0$.
 Based on the assumption of $\epsilon$ \eqref{eq41p}, it follows that $(1+s) \mu_{B} \epsilon/(1-c_{\phi s}) 
 \leq \underline{\mu_x}/{6}$.
According to \eqref{eq60p}, when $k={K}_0+1$, for  $ \forall \,i \in S$, we have
\begin{align} \label{eq78}
	\| \tilde{\bm{X}}_{i,:}^{{K}_0+1} - \tilde{\bm{X}}_{i,:}^* \|_2
	 \leq
	\| \tilde{\bm{X}}^{{K}_0+1}  - \tilde{ \bm{X}}^* \|_{F}
	\leq C_0 (c_{\phi s})^{{K}_0+1} 
	+ \frac{(1+s) \mu_{B} \epsilon}{1-c_{\phi s}} 
	< 
	\frac{\underline{\mu_x}}{3}
	< 
	\underline{\mu_x}.
\end{align}

Then, we  prove that  $\| \tilde{\bm{X}}_{i,:}^{{K}_0+1} \|_2>0, \forall\,i\in S$.
If $\| \tilde{\bm{X}}_{i,:}^{{K}_0+1} \|_2=0$,
then we have $\tilde{\bm{X}}_{i,j}^{{K}_0+1}=0, \forall j\in [M]$.
Hence, we obtain $\| \tilde{\bm{X}}_{i,:}^{{K}_0+1} - \tilde{\bm{X}}_{i,:}^* \|_2 = \| \tilde{\bm{X}}_{i,:}^* \|_2 <\underline{\mu_x}$,
which contradicts with the assumption
$\| \tilde{ \bm{X}}^*_{i,:} \|_2\geq\underline{\mu_x}>0$.
Therefore, we obtain  $\| \tilde{\bm{X}}_{i,:}^{{K}_0+1} \|_2>0$.

The definition of $\eta_k$ in \eqref{eq45p} implies that $\eta_k=\frac{1}{2\theta_k}$ when $k\geq{K}_0$.
Hence, the univariate proximal operator $\hat{P}_{\theta_k,f_{\eta_k}}(\cdot)$  becomes  a hard thresholding function, i.e.,
\begin{align}\label{eq76p}
 \hat{P}_{\theta_k,f_{\eta_k}}(x) = \left \{
	\begin{aligned}
		& 0, 
		&& \text{if}\,|x| \leq \theta_k , \\
		& x,      
		&&\text{if}\, |x| > \theta_k.
	\end{aligned}
	\right. 
\end{align}
With \eqref{eq56p}, \eqref{eq76p}, and $\| \tilde{\bm{X}}_{i,:}^{{K}_0+1} \|_2>0$, we  obtain $\partial g_{\eta_{{{K}_0}}}(\|\tilde{\bm{X}}_{i,:}^{{{K}_0+1}}\|_2)=\bm{0}$.
Hence, we prove  that $\partial g_{\eta_{{{K}_0}}}(\|\tilde{\bm{X}}_{S,:}^{{{K}_0+1}}\|_2)=\bm{0}$.

(ii) We assume that $\partial g_{\eta_{t}}(\|\tilde{\bm{X}}_{S,:}^{t+1}\|_2)=\bm{0}$ for $ {K}_0\leq t\leq k$.
According to \eqref{eq68p} and $\bm{M} = \bm{T}\bm{\Lambda_{\bm{M}}}\bm{T}^{-1}$, we have
\begin{align}
	\bm{z}^{t} = 
	\bm{T}\bm{\Lambda_{\bm{M}}}\bm{T}^{-1} \bm{z}^{t-1}, \quad {K}_0+1\leq t\leq k.
\end{align}
Then, we obtain
\begin{align}\label{eq73p}
	\bm{z}^{k} = 
	\bm{T}(\bm{\Lambda_{\bm{M}}})^{k-{K}_0}\bm{T}^{-1} \bm{z}^{{K}_0}.
\end{align}
By taking norm on both sides of \eqref{eq73p}, we have
\begin{align}\label{eq74p}
	\|\bm{z}^{k}\|_F 
	\leq  
	\|\bm{T}\|_F\|\bm{\Lambda_{\bm{M}}}\|_F^{k-{K}_0}\|\bm{T}^{-1}\|_F \|\bm{z}^{{K}_0}\|_F
	\leq \frac{2s(1+\hat{\beta})\|\bm{z}^{{K}_0}\|_F}{\sqrt{4\hat{\beta}-(\hat{\beta}+\phi s -\phi)^2}}\bigg(\sqrt{2s\hat{\beta}}\bigg)^{k-{K}_0}.
\end{align}

Next, we bound $\|\bm{z}^{{K}_0}\|_F$.
Based on \eqref{eq64p}, $ \beta_{{K}_0}=0$ and $\partial g_{\eta_{{{K}_0}}}(\|\tilde{\bm{X}}_{S,:}^{{{K}_0+1}}\|_2)=\bm{0}$, we have
\begin{align}\label{eq75p}
	\tilde{\bm{X}}^{{K}_0+1}_{S,:}-\bar{\bm{X}}_{S,:}
	= (\bm{I}_S - \bm{B}_{S,:}\tilde{\bm{S}}_{:,S})(\tilde{\bm{X}}^{{K}_0}_{S,:}-\bar{\bm{X}}_{S,:}).
\end{align}
Due to the definition of $\phi$, we know that the elements of matrix $\bm{B}_{S,:}\tilde{\bm{S}}_{:,S}$ except the diagonal elements are not larger than $\phi$, and the diagonal elements are zero.
Thus, we have $\| \bm{I}_S - \bm{B}_{S,:}\tilde{\bm{S}}_{:,S}\|_F
 \leq \sqrt{|S|(|S|-1)}\phi \leq \sqrt{s(s-1)}\phi$.

By taking norm on both sides of \eqref{eq75p}, we have
\begin{align}
	\|\tilde{\bm{X}}^{{K}_0+1}_{S,:}-\bar{\bm{X}}_{S,:}\|_F
	&\leq
	\| \bm{I}_S - \bm{B}_{S,:}\tilde{\bm{S}}_{:,S}\|_F
	\| \tilde{\bm{X}}^{{K}_0}_{S,:}-\bar{\bm{X}}_{S,:}
	\|_F \notag
	\\&\leq
	 \sqrt{s(s-1)}\phi 
	\| \tilde{\bm{X}}^{{K}_0}_{S,:}-\bar{\bm{X}}_{S,:}
	\|_F \notag
	\\&\leq(2s-1)\phi
	\| \tilde{\bm{X}}^{{K}_0}_{S,:}-\bar{\bm{X}}_{S,:}
	\|_F 
	\leq
	\| \tilde{\bm{X}}^{{K}_0}_{S,:}-\bar{\bm{X}}_{S,:}
	\|_F.
\end{align}
Recalling that $\bm{z}^{{K}_0} = [(\tilde{\bm{X}}^{{K}_0+1}_{S,:}-\bar{\bm{X}}_{S,:})^{\text{T}},(\tilde{\bm{X}}^{{K}_0}_{S,:}-\bar{\bm{X}}_{S,:})^{\text{T}}]^{\text{T}}$, we have 
\begin{align}\label{eq77p}
	&\|\bm{z}^{{K}_0}\|_F 
	\leq 2\|\tilde{\bm{X}}_{S,:}^{{K}_0}-\bar{\bm{X}}_{S,:}\|_F
	\leq 2\|\tilde{\bm{X}}_{S,:}^{{K}_0}-\tilde{\bm{X}}_{S,:}^*\|_F
	+ 2\|\delta\tilde{\bm{X}}_{S,:}\|_F \notag
	\\&\leq 
	2\sqrt{s}\mu_x (c_{\phi s})^{{K}_0} 
	+ \frac{2(1+s) \mu_{B} \epsilon}{1-c_{\phi s}}
	+ \frac{2\sqrt{s} \mu_{B}}{1+\phi-\phi s}
	\epsilon \notag
	\\&\leq 
	2\sqrt{s}\mu_x (c_{\phi s})^{{K}_0}  
	+ \frac{2(1+s+\sqrt{s}) \mu_{B}\epsilon}{1-c_{\phi s}}\notag
	\\&
	\leq 2\sqrt{s}\mu_x (c_{\phi s})^{{K}_0} 
	+ 2\sqrt{s}\mu_x (c_{\phi s})^{{K}_0}  
	= 
	4\sqrt{s}\mu_x (c_{\phi s})^{{K}_0},
\end{align}
where the last inequality follows from \eqref{eq41p}.


Combining with \eqref{eq74p}, we obtain
\begin{align} \label{eq78p}		  &\|\tilde{\bm{X}}^{k+1}_{S,:}-\bar{\bm{X}}_{S,:}\|_F 
	\!\leq\!
	\|\bm{z}^k\|_F
	\!\leq \!
	\frac{8\mu_x s \sqrt{s} (1+\hat{\beta})(c_{\phi s})^{{K}_0} }{\sqrt{4\hat{\beta}-(\hat{\beta}+\phi s -\phi)^2}}\bigg(\sqrt{2s\hat{\beta}}\bigg)^{k-{K}_0}
	\!\leq\!
	C_0(c_{\phi s})^{{K}_0+1}
	\bigg(\sqrt{2s\hat{\beta}}\bigg)^{k-{K}_0}.
\end{align}


Subsequently, by defining 
$
	\hat{\bm{X}}^{k+1}_{S,:} = \tilde{\bm{X}}^{k+1}_{S,:}  - 
	\bm{B}_{S,:}\tilde{\bm{S}}_{:,S} 
	(\tilde{\bm{X}}^{k+1}_{S,:}-\bar{\bm{X}}_{S,:}) 
	+ \beta_{k+1} (\tilde{\bm{X}}^{k+1}_{S,:}-\tilde{\bm{X}}^{k}_{S,:})
$
and
$\hat{\bm{z}}^k = [(\hat{\bm{X}}^{k+1}_{S,:}-\bar{\bm{X}}_{S,:})^{\text{T}},(\tilde{\bm{X}}^{k+1}_{S,:}-\bar{\bm{X}}_{S,:})^{\text{T}}]^{\text{T}}
$, 
we have
$
	\hat{\bm{z}}^k = \bm{M}\bm{z}^k $
	and
	$
	\tilde{\bm{X}}^{k+2}_{S,:} = P_{\theta_{k+1},f_{\eta_{k+1}}}(\hat{\bm{X}}^{k+1}_{S,:})$.
Following the same idea of proving \eqref{eq78p}, we obtain
\begin{align} 				    \|\hat{\bm{X}}^{k+1}_{S,:}-\bar{\bm{X}}_{S,:}\|_F 
	\leq
	\|\hat{\bm{z}}^k\|_F
	\leq
	C_0(c_{\phi s})^{{K}_0+1}
	\bigg(\sqrt{2s\hat{\beta}}\bigg)^{k+1-{K}_0}.
\end{align}
Since the inequality $\sqrt{2s\hat{\beta}}
=
1-\sqrt{1-c_{\phi s}}
\leq 
c_{\phi s}$ holds when $c_{\phi s} < 1$, we have
\begin{align}
	\|\hat{\bm{X}}^{k+1}_{S,:}-\bar{\bm{X}}_{S,:}\|_F 
	\leq
	C_0(c_{\phi s})^{k+2}
	<
    \underline{\mu_x}/6.
\end{align}
Recalling that
$\bar{\bm{X}}_{S,:} = \tilde{\bm{X}}_{S,:}^*+\delta\tilde{\bm{X}}_{S,:}$, for  $ \forall \,i\in S$,
we obtain 
\begin{align}
	\|\hat{\bm{X}}^{k+1}_{i,:}-\tilde{\bm{X}}^*_{i,:}\|_2
	\leq
	\|\hat{\bm{X}}^{k+1}_{S,:}-\tilde{\bm{X}}^*_{S,:}\|_F
	\leq
	\|\hat{\bm{X}}^{k+1}_{S,:}-\bar{\bm{X}}_{S,:}\|_F
	+ \|\delta\tilde{ \bm{X}}_{S,:}\|_F
	< 
	\underline{\mu_x}/6 + \underline{\mu_x}/6
	= \underline{\mu_x}/3.
\end{align}
Hence, for  $\forall \,i\in S$, we have
\begin{align}
	\| \hat{ \bm{X}}^{k+1}_{i,:} \|_2
	\geq
	\| \tilde{ \bm{X}}^*_{i,:} \|_2
	-
	\| \hat{ \bm{X}}^{k+1}_{i,:} - \tilde{ \bm{X}}^*_{i,:} \|_2
	>
	\underline{\mu_x}
	-
	\underline{\mu_x}/3
	= 2\underline{\mu_x}/3.
\end{align}
With \eqref{eq65p}, we obtain
\begin{align}
	\| \delta\tilde{\bm{X}}_{S,:} \|_F
	\leq 
	\frac{\sqrt{s} \mu_{B}\epsilon}{1+\phi-\phi s}
	\leq 
	\frac{\sqrt{s} \mu_{B}\epsilon}{1+\phi-2\phi s}
		\leq
		\frac{\underline{\mu_x}}{6}
		\frac{\sqrt{s}}{s+1}
		\leq
		\frac{\underline{\mu_x}}{6}.
\end{align}
By the definition of $\hat{\beta}$, we have
\begin{align}
	\| \tilde{\bm{X}}^{k+1}_{i,:}  - \bar{ \bm{X}}^*_{i,:} \|_{2}
	\leq
	\|\tilde{\bm{X}}^{k+1}_{S,:}-\bar{\bm{X}}_{S,:}\|_F 
	\leq
	C_0(c_{\phi s})^{{K}_0+1}
	\bigg(\sqrt{2s\hat{\beta}}\bigg)^{k-{K}_0}
	\leq
	C_0(c_{\phi s})^{k+1}
	\leq
	\underline{\mu_x}/6.
\end{align}
Thus, we obtain
\begin{align}
	\| \tilde{\bm{X}}^{k+1}_{i,:}  -  \tilde{\bm{X}}^*_{i,:} \|_{2}
	\leq 
	\| \tilde{\bm{X}}^{k+1}_{i,:}  - \bar{ \bm{X}}^*_{i,:} \|_{2}
	+ 
	\|\delta \tilde{\bm{X}}^{k+1}_{i,:} \|_2
	\leq \underline{\mu_x}/6 +  \underline{\mu_x}/6 = \underline{\mu_x}/3.
\end{align}

On the other hand, according to the definition of $\theta_{k}$, we have
\begin{align} 
	\theta_{k+1} 
	&=   \phi \mathop{\text{sup}}_{(\tilde{\bm{X}}^*,\tilde{\bm{Z}})\in\mathcal{X}(\underline{\mu_x},\mu_x,s,\epsilon)  }  
	\| \tilde{\bm{X}}^{k+1}  - \tilde{ \bm{X}}^* \|_{2,1} 
	+ \mu_{B} \epsilon \notag
	\\&=  \phi \mathop{\text{sup}}_{(\tilde{\bm{X}}^*,\tilde{\bm{Z}})\in\mathcal{X}(\underline{\mu_x},\mu_x,s,\epsilon)  }  
	\sum_{i=1}^{|S|} \| \tilde{\bm{X}}^{k+1}_{i,:}   - \tilde{\bm{X}}^*_{i,:}  \|_{2} 
	+ \mu_{B} \epsilon \notag
	\\&\leq \phi s  \underline{\mu_x}/3 + \mu_{B} \epsilon 
	\leq \underline{\mu_x}/3 + \underline{\mu_x}/3 
	 = 2\underline{\mu_x}/3.
\end{align}

Because of $\tilde{\bm{X}}^{k+2}_{S,:} = P_{\theta_{k+1},f_{\eta_{k+1}}}(\hat{\bm{X}}^{k+1}_{S,:})$ and the inequality $\| \hat{ \bm{X}}^{k+1}_{i,:} \|_2>2\underline{\mu_x}/3\geq\theta_{k+1}$,
we obtain $\tilde{\bm{X}}^{k+2}_{i,:}=\hat{\bm{X}}^{k+1}_{i,:}, \forall\, i\in S$.
Thus,  we prove that $\| \tilde{ \bm{X}}^{k+2}_{i,:} \|_2>0$.
Finally, we can obtain $\partial g_{\eta_{k+1}}(\|\tilde{\bm{X}}_{S,:}^{k+2}\|_2)=\bm{0}$.

We have  proved $\partial g_{\eta_t}(\|\tilde{\bm{X}}_{S,:}^{t+1}\|_2)=\bm{0}$  for all $t$ satisfying ${K}_0 \leq t \leq k \,(k\geq{K}_0)$ by induction.
According to \eqref{eq78p}, we have
\begin{align}
	&\|\tilde{\bm{X}}^{k+1}-\tilde{\bm{X}}^*\|_F
	\leq
	\|\tilde{\bm{X}}^{k+1}_{S,:}-\bar{\bm{X}}_{S,:}\|_F
	+ \|\delta\tilde{\bm{X}}_{S,:} \|_F \notag
	\\&\leq
	C_0(c_{\phi s})^{{K}_0+1}
	\bigg(1-\sqrt{1-c_{\phi s}}\bigg)^{k-{K}_0}
	+ 
	\frac{s \mu_{B}}{1+\phi-\phi s}
	\epsilon,\quad \forall\, k\geq K_0 +1.
\end{align}

%
%

\bibliographystyle{IEEEtran}
\bibliography{ref}

\end{document}